\documentstyle[pre,preprint,aps]{revtex} 
\input epsf
\newcounter{abc}

\newcommand{\gE}{\mbox{$\tau$}} 
\newcommand{\gM}{\mbox{$h$}}
\newcommand{\oE}{\mbox{${\cal E}$}} 
\newcommand{\oM}{\mbox{${\cal M }$}}
\newcommand{\aE}{\mbox{$a_{{\cal E}}$}} 
\newcommand{\aM}{\mbox{$a_{{\cal M}}$}}
\newcommand{\LM}{\mbox{$\Lambda _{{\cal M}}$}}
\newcommand{\LMp}{\mbox{$\Lambda _{\cal M}^+$}}
\newcommand{\LE}{\mbox{$\Lambda _{{\cal E}}$}}
\newcommand{\LEp}{\mbox{$\Lambda _{{\cal E}}^+$}}
\newcommand{\ptME}{\mbox{$\tilde{p}_{{\cal M},{\cal E}}$}}

\newcommand{\ptMstar}{\mbox{$\tilde{p}^\star_{\cal M}(x)$}}
\newcommand{\ptMEstar}{\mbox{$\tilde{p}^\star_{{\cal M},{\cal E}}$}}

\newcommand{\pLM}{\mbox{$p_L({\cal M})$}} 
\newcommand{\pLME}{\mbox{$p_L({\cal M},{\cal E})$}} 
\newcommand{\pLE}{\mbox{$p_L({\cal E})$}} 
\newcommand{\mix}{\mbox{$s$}}
\newcommand{\mixp}{\mbox{$r$}}
\newcommand{\dfac}{\mbox{$\frac{1}{1-\mix \mixp}$}}
\begin{document}

\title{Simulation studies of fluid critical behaviour}

\author{Nigel B. Wilding}
\address{ Institut f\"{u}r Physik, Johannes Gutenberg-Universit\"{a}t Mainz,\\
Staudinger Weg 7, D-55099 Mainz, Germany.}

\maketitle
\tighten
\begin{abstract}

We review and discuss recent advances in the simulation of bulk critical
phenomena in model fluids.  In particular we emphasise the extensions to
finite-size scaling theory needed to cope with the lack of symmetry
between coexisting fluid phases.  The consequences of this asymmetry for
simulation measurements of quantities such as the particle density and
the heat capacity are pointed out and the relationship to experiment is
discussed.  A general simulation strategy based on the finite-size
scaling theory is described and its utility illustrated via Monte-Carlo
studies of the Lennard-Jones fluid and a two-dimensional spin fluid
model.  Recent applications to critical polymer blends and solutions are
also briefly reviewed.  Finally we consider the outlook for future
simulation work in the field. 

\end{abstract}
\newpage

\section{Introduction}

That the vapour pressure curve of a simple fluid terminates in a critical
point has been known since the latter half of the last century when Andrews
demonstrated the phenomenon of critical opalesence in carbon dioxide
\cite{ANDREWS}.  Shortly afterwards, van der Waals published a mean field
type theory of liquid-vapour coexistence that predicted the existence of a
critical point with divergent compressibility, and formed much of the basis
of the understanding of the critical region for many years thereafter. 
Arguably, however, the modern era of critical phenomena began in the 1940s,
when Guggenheim realised that the coexistence curves of many simple fluids
are not in fact parabolic as predict by van der Waals theory, but nearly
cubic.  At around the same time, Onsager published his famous solution to
the 2D ferromagnetic Ising model, which among other things allowed it to be
demonstrated that the critical exponents are strongly non-classical.  It was
not until the 1960s, however, that activity in the field really began to
take off.  Accurate experimental studies of magnetic systems on the one
hand, and series expansion studies of model spin systems on the other,
providing vivid illustration of the phenomena of scaling and universality,
already hinted at previously by Guggenheim for fluids \cite{STANLEY}.  These
findings heralded the beginning of a prodigious effort to understand more
deeply the nature of critical phenomena, and in particular of finding ways
of dealing with the profusion of length scales that characterise the
critical region and which underpin the singularities in physical observables
that are its hallmark.  Milestones in this quest to date, are the
renormalisation group \cite{BINNEY} and conformal invariance theories
\cite{HENKEL}, the introduction of which have added substantially to the
armoury of the critical point theorist. 

Notwithstanding the great strides made in the understanding of critical
phenomena over recent years, it is the exception rather than the rule that
the available theoretical apparatus permits exact calculation of the
critical properties of a given model system.  Many of the systems of
interest to physicists have steadfastly resisted efforts to find accurate
analytical treatments of their critical properties.  Indispensable,
therefore, in tackling these analytically recalcitrant models has been
computer simulation.  Methods such as finite-size scaling or the Monte-Carlo
renormalisation group permit one, in principle, to extract accurate
estimates for bulk critical properties from simulations of finite-sized
systems.  Until recently, however, such simulation studies of critical
phenomena were principally confined to simple lattice-based magnetic systems
like the $O(N)$ or q-state Potts models.  More complex systems such as
off-lattice fluids were deemed to be simply too computationally demanding. 
Gradually, though, and with the advent of new simulation methodologies, and
ever more powerful computers, this situation is changing and it is now
becoming possible to perform detailed simulation investigations of critical
phenomena in a variety of simple and complex fluid systems.  These
developments are opening up for simulation study, a whole new vista of
interesting critical and phase coexistence behaviour.  Phenomena of topical
interest encompass not only the standard liquid-vapour or liquid-liquid
(consolute) critical phenomena, but also multicritical and crossover
behaviour in diverse classes of systems ranging from simple atomic fluids to
electrolytes and polymer blends. 
   
In this article we review recent progress in the simulation of bulk
criticality in model fluid systems.  We begin by describing recent advances
in finite-size scaling (FSS) theory for fluids.  These advances generalise
the FSS techniques previously developed in the magnetic context in order to
take account of the `field mixing' phenomenon that manifests the lack of
symmetry between coexisting fluid phases.  This broken symmetry, which is a
fundamental issue in the critical behaviour of fluids, is shown to have
important consequences for simulation measurements of quantities such as the
particle density, or the specific heat.  The relationship to experimental
studies of field mixing is also discussed.  We then describe a simulation
methodology based on the insights derived from the FSS theory, and
illustrate its utility by describing, in some detail, recent studies of
critical phenomena in the Lennard-Jones fluid and tricritical phenomena in a
two-dimensional spin fluid model.  Attention is also drawn to recent FSS
studies of critical behaviour in polymer blends and solutions.  Finally we
assess the prospects for further work in the field and highlight a number of
outstanding unresolved issues. 

\section{Finite-size scaling theory for near-critical fluids}

\label{sec:back}

In the neighbourhood of a critical point, the correlation length $\xi$ grows
extremely large and often exceeds the linear size $L$ of the simulated system. 
When this occurs, the singularities and discontinuities that characterise
critical phenomena in the thermodynamic limit are smeared out and shifted
\cite{BINDER1}.  Unless care is exercised, such finite-size effects can lead
to serious errors in computer simulation estimates of critical point
parameters. 

To deal with these problems, finite-size scaling (FSS) techniques have been
developed \cite{PRIVMAN}.  Use of FSS methods enable one to extract accurate
estimates of infinite-volume quantities from simulations of finite-size.  As
its basic tenet, the FSS hypothesis holds that for sufficiently large $\xi$
and $L$, the coarse-grained properties of a given near-critical system are
universal \cite{KADANOFF} and depend (up to certain non-universal factors)
only on specific combinations of $L$ and the relevant scaling fields that
measure deviations from criticality \cite{WEGNER}.  Systems having
short-ranged interactions and a single component order-parameter (such as
many fluids and magnets) belong to the Ising universality class
\cite{SENGERS,BACK}, ie.  their scaling functions and critical exponents are
identical to those of the simple Ising model.  However, qualitatively
different systems, such as those with multi-component order-parameters may
exhibit quite different universal behaviour. 

For the purpose of setting out the general FSS concepts, however, we shall
remain with the Ising universality class, for which the critical behaviour
is controlled by two relevant scaling fields, which we denote \gE\ and \gM. 
For the Ising model itself, (or equivalently, the ordinary lattice gas)
there is a special `particle-hole' symmetry with respect to change of sign
of the magnetic field $H$.  This symmetry implies that fluctuations in the
order-parameter $\delta m=m-m_c$, and the energy density $\delta u=u-u_c$
are orthogonal (statistically independent), ie.  $\langle \delta m \delta
u\rangle=0$.  As a result, the phase boundary of the Ising model coincides
with the $H=0$ axis of the phase diagram and thus it is natural to assign
$\gE = T-T_c$ and $\gM=H-H_c$ \cite{PRIVMAN}. 

Conjugate to the scaling fields \gE\ and \gM\ are scaling operators \oE\ and
\oM .  For the Ising model, one has $\oE= u$ (the energy density) and $\oM =
m$ (the magnetisation).  In the near-critical region both \oE\ and \oM\
fluctuate strongly and thus their coarse grained (large length scale)
properties are expected to exhibit universal scaling behaviour.  General finite-size
scaling arguments \cite{BRUCE2,WILDING1,BINDER2,BRUCE1,PRIVMAN} predict that
the joint distribution \pLME\ exhibits scaling behaviour of the form

\setcounter{abc}{1}
\begin{equation}
\label{eq:ansatz}
p_L(\oM , \oE) \simeq  \LMp \LEp \ptME (\LMp \delta \oM , \LEp \delta \oE , \LM \gM , \LE \gE )
\end{equation}
\addtocounter{equation}{-1}
\addtocounter{abc}{1}
where 
\begin{equation}
\label{eq:Lamdefs}
\LE = \aE L^{1/\nu} \hspace{1cm} \LM = \aM L^{d-\beta/\nu}\hspace{1cm} \LM \LMp = \LE \LEp = L^d 
\end{equation}
\addtocounter{equation}{-1}
\addtocounter{abc}{1}
and
\begin{equation}
\label{eq:deltops}
\delta \oM \equiv  \oM - <\oM >_c  \hspace{1cm} \delta \oE \equiv  \oE - <\oE >_c 
\end{equation}
\setcounter{abc}{0}
where $d$ is the system dimensionality and $\nu$ and $\beta$ are the
standard critical exponents for the correlation length and order-parameter
respectively \cite{STANLEY}.  The subscripts c in equations~\ref{eq:deltops}
signify that the averages are to be taken at criticality. 

The forms of the $L$ dependences of the operator distributions in
equation~\ref{eq:Lamdefs} arise from the relationship between the
correlation length $\xi$, and the system size.  For example, in the case of
the ordering operator, one has $\delta {\cal M} \sim
(\tau_c-\tau)^{\beta}\sim\xi^{-\beta/\nu}$.  Since, however, at criticality,
the correlation length is bounded by $L$, one can simply substitute $\xi\to
L$, giving $\delta{\cal M}\sim L^{-\beta/\nu}$.  A similar arguement
pertains to ${\cal E}$.  To obtain the dependence of the operator
distributions on the scaling fields, one assumes that the scaling behaviour
depends solely on the ratio $L/\xi$ \cite{PRIVMAN}.  The known dependences
of $\xi$ on $\tau$ and $h$ then yield the given combinations. 

Given appropriate choices for the non-universal scale factors \aM\ and \aE\
(equation~\ref{eq:Lamdefs}), the function \ptME\ is expected to be
universal.  Precisely at criticality, the scaling fields \gE\ and \gM\
vanish by definition, implying that

\begin{equation}
p_L(\oM , \oE) \simeq \LMp \LEp \ptMEstar (\LMp \delta \oM , \LEp \delta \oE),
\label{eq:critlim}
\end{equation}
where $\ptMEstar(x,y)\equiv\ptME(x,y,0,0)$ is a function
describing the universal and statistically scale invariant operator
fluctuations characteristic of the critical point. It follows that $\ptMEstar(x,y)$ 
constitutes a {\em hallmark} of a universality class, a fact that (as we
shall see) can often be exploited to obtain accurate estimates of the critical point
parameters of fluid systems.

\begin{figure}[h]
\setlength{\epsfxsize}{13cm}
\centerline{\mbox{\epsffile{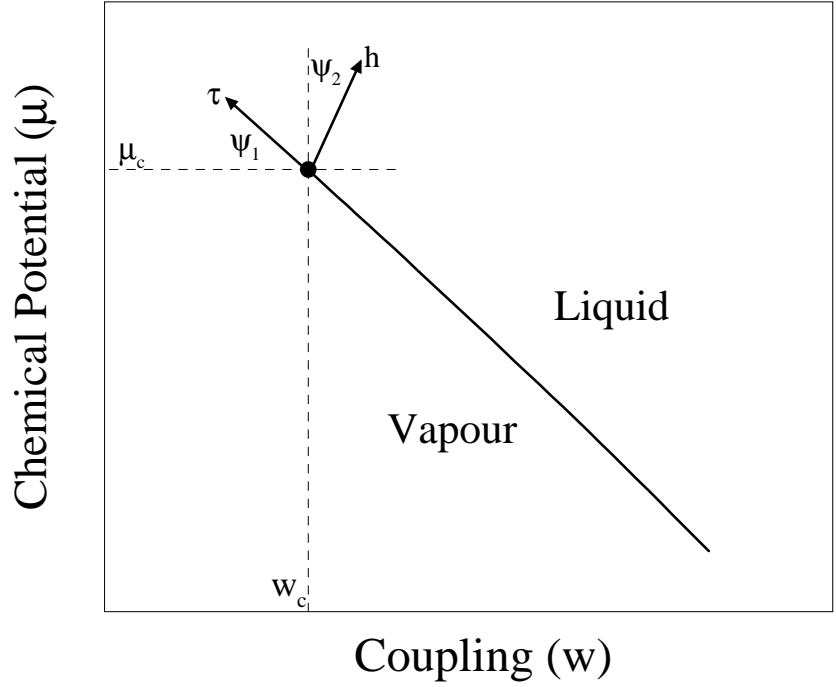}}} 

\caption{Schematic representation of the liquid-vapour coexistence
curve showing the directions of the relevant scaling scaling
fields. The angles $\psi_1$ and $\psi _2$ are related to the
field-mixing parameters \mix\ and \mixp\ (equation
\protect\ref{eq:sfdefs}) by $ \mixp = -\tan \psi _1 $ and $ \mix =
\tan \psi _2 $.}

\label{fig:scaflds}
\end{figure}

Although the generality of the FSS expression, equation~\ref{eq:ansatz},
encompasses all members of the Ising universality class, irrespective of the
symmetry between the coexisting phases, it has long been appreciated
\cite{REHR} that the {\em form} of the fluid scaling fields differ from
those of the Ising model.  Specifically, the reduced symmetry of fluids (ie. 
the fact that $\langle \delta\rho\delta u\rangle\neq 0$, with $\rho$ the
density) leads to `mixed' scaling fields comprising {\em
linear-combinations} of the reduced coupling\footnote{For convenience we
will adopt the reduced coupling $w=J/k_BT$ in this definition of the scaling
fields, rather than the temperature itself.  Both definitions are
permissible.}(inverse temperature) $w$ and applied (reduced chemical
potential) field $\mu$:

\begin{equation} 
\tau = w_c-w+s(\mu - \mu_c) \hspace{1cm}
h=\mu - \mu_c+ r(w_c-w), 
\label{eq:sfdefs} 
\end{equation} 
where the parameters \mix\ and \mixp\ are non-universal (system-specific) quantities
controlling the degree of field mixing. In particular, $\mixp$ is
identifiable as the limiting critical gradient of the coexistence
curve in the space of $\mu$ and $w$. The role of $s$ is somewhat
less tangible; it controls the degree to which the chemical potential
features in the weak scaling field $\tau$. The directions of the fluid scaling fields 
are indicated schematically in the phase diagram of figure~\ref{fig:scaflds}.

As a result of the mixed character of the fluid scaling fields $\tau$ and
$h$, the respective conjugate scaling operators ${\cal E}$ and ${\cal M}$
are found to comprise linear combinations \cite{BRUCE2,WILDING1} of the
order-parameter (particle density $\rho$) and the energy density $u$. 
Specifically, one finds

\begin{equation} 
\oM  = \dfac \left[ \rho - s u \right], \hspace{1cm} \oE  =  \dfac \left[  u  - r \rho \right] .
\label{eq:oplinks} 
\end{equation} 
We shall henceforth term \oM\ the ordering operator and \oE\ the energylike operator.
For the correct choice of the field mixing parameters $s$ and $r$, the
operators satisfy the relation $\langle \delta\oM\delta \oE \rangle= 0$. 
The forms of the scaling fields and scaling operators for single component
systems with and without field mixing are summarised in
table~\ref{table:relns}. 

The operators \oM\ and \oE\ are the generalised scaling counterparts of the
Ising magnetisation and energy.  Typically, however, in fluid simulation
studies, one is more concerned with obtaining estimates for the critical
density and energy density.  Owing to field mixing, these quantities are not
expected to exhibit the same FSS behaviour as the Ising magnetisation or
energy.  To elucidate their properties it is expedient to reexpress $\rho$
and $u$ in terms of the scaling operators.  Appealing to
equation~\ref{eq:oplinks}, one finds

\begin{equation}
u=\oE-\mixp\oM , \hspace{1cm} \rho=\oM - s \oE ,
\label{eq:udenmix}
\end{equation}
so that the critical density and energy density distributions are

\begin{equation}
p_L(u)=p_L (\oE -r\oM ) \hspace{1cm} p_L(\rho)=p_L (\oM -s\oE ) .
\label{eq:prho&u}
\end{equation}

Now the structure of the scaling form~\ref{eq:critlim} shows that the
typical size of the fluctuations in the energy-like operator will vary with
system size like $\delta\oE\sim L^{-(d-1/\nu)}\sim L^{-(1-\alpha)/\nu}$, where
$\alpha$ is the specific heat exponent and we have employed the hyperscaling
relation $d\nu=2-\alpha$.  The typical size of the fluctuations in the
ordering operator, on the other hand, vary like $\delta\oM\sim
L^{-\beta/\nu}$.  From this it is is easy to show \cite{WILDING3} that for
a given $L$, the shape of the energy and density distributions can be
identified with the distribution of the variable

\begin{equation}
X_{\Theta} = a_{\cal M}^{-1}\delta \oM \cos \Theta
+ a_{\cal E}^{-1}\delta\oE \sin \Theta  ,
\label{eq:Theta}
\end{equation}
with

\begin{equation} 
\tan \Theta_{u} = \frac{a_{\oE}}{r a_{\oM}} L^{-(1-\alpha -\beta)/\nu}
\hspace{0.5cm} , \hspace{0.5cm}\tan \Theta_{\rho} = \frac{s a_{\oE}}{a_{\oM}}
L^{-(1-\alpha -\beta)/\nu} 
\label{eq:Ldep}
\end{equation}
where the subscripts $u$ and $\rho$ signify that the value of $\Theta$
corresponds to the energy density and particle density distributions
respectively. 

The distributions $p(X_\Theta)$ constitute a spectrum of {\em universal}
functions (parameterised by the value of $\Theta$) describing the particle
density and energy distributions of fluids at finite $L$
\cite{WILDING3,WILDING2}.  This $L$ dependence arises from the different
relative strengths of the critical fluctuations in \oM\ and \oE .  Since
$1-\alpha > \beta$, the critical fluctuations in \oM\ are stronger than those
in \oE , causing the distribution \pLE\ to converge to its average value more
rapidly with increasing $L$, than \pLM .  Consequently, measurements of the
distributions of $\rho$ and $u$, (each of which represent linear combinations
of \oM\ and \oE ), yield $L$-dependent functional forms.

\begin{figure}[t]
\setlength{\epsfxsize}{14cm}
\centerline{\mbox{\epsffile{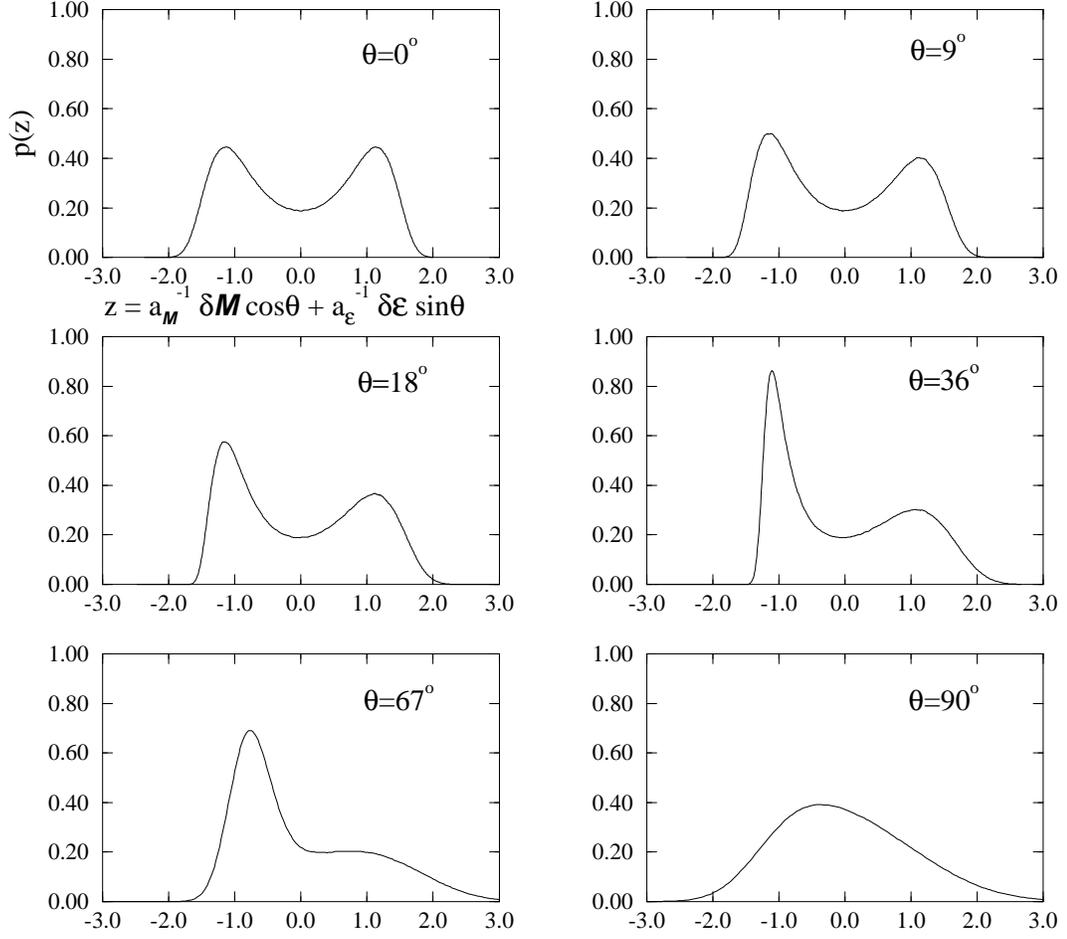}}} 

\caption{Selections from the universal finite-size spectrum of critical
density and energy density distributions of fluids.  The distributions were
obtained according to the procedure described in the text.  Following
convention, the values of the non-universal scale factors $a_{\cal E}^{-1}$
and $a_{\cal M}^{-1}$ have been chosen to ensure that the distributions have
unit variance.  From ref.  \protect\cite{WILDING3}}

\label{fig:projs}
\end{figure}

Figure ~\ref{fig:projs} shows the form of $p(X_\Theta)$ for a representative
selection of values of $\Theta$.  The distributions were constructed from
the joint distribution $p_L(m,u)$ of the critical Ising model obtained in
reference \cite{WILDING3}.  Since for the Ising model, ${\cal M}\to m$ and ${\cal
E}\to u$, the form of $p(X_\Theta)$ is obtained simply by taking linear
combinations of the form $p_L(\delta m\cos\Theta +\delta u\sin \Theta )$. 
For $\Theta=0^\circ$ this yields, trivially, the ordering operator
distribution $p_L({\cal M})$, while for $\Theta=90^\circ$ the form is that
of the energylike operator distribution $p_L({\cal E})$.  Intermediate
between these values a range of behaviour is obtained, representing the
finite $L$ forms of $p_L(\rho )$ and $p_L(u)$ for fluids. Of course,
since $p_L({\cal M},{\cal E})$ is universal, then so also
are all the finite-size forms of $p_L(\rho )$ and $p_L(u)$. 

In the limit $L\rightarrow \infty$, equation~\ref{eq:Ldep}
implies that both $\Theta_u$ and $\Theta_\rho$ approach zero
so that
\begin{eqnarray}
\setcounter{abc}{1}
p_L(u)&=& p_L (-r\oM )\simeq\aM ^{-1}\mixp L^{\beta/\nu
}\tilde{p}_{\cal M}^\star (-\aM ^{-1}\mixp L^{\beta/\nu} \delta \oM )\\
\label{eq:lim_u}
\addtocounter{equation}{-1}
\addtocounter{abc}{1}
p_L(\rho)&=& p_L (\oM )\simeq\aM ^{-1} L^{\beta/\nu
}\tilde{p}_{\cal M}^\star (\aM ^{-1} L^{\beta/\nu} \delta \oM )
\label{eq:lim_rho}
\end{eqnarray}
\setcounter{abc}{0}
It follows that for {\em any} finite \mix\ and \mixp , the limiting critical
point forms of $p_L(\rho)$ and $p_L(u)$ both match the critical ordering
operator distribution \ptMstar=$\int dy$\ptMEstar\ $(x,y)$ . This result
reflects the fact that for sufficiently large $L$, the critical fluctuations
in \oE\ are negligible on the scale of those in \oM . It should be noted,
however, that a quite different state of affairs obtains for the critical
Ising model where, owing to the absence of field mixing ($s=r=0$),
$\lim_{L\to\infty}p_L(u)= p_L({\cal E })$.

This alteration to the asymptotic behaviour of $p_L(u)$ turns out to have
important ramifications for the manner in which the specific heat is
measured.  Within the canonical ensemble of the Ising model, the specific
heat is most readily calculated from the variance of the energy
fluctuations, which at criticality scales with system size like

\begin{equation}
C_v=L^d(\langle u^2\rangle-\langle u \rangle^2)/k_BT^2\propto L^{\alpha/\nu}.
\end{equation}
For fluids, however, this is not the correct definition to use
since the alteration to the limiting form of $p_L(u)$ implies that 
\begin{equation}
L^d(\langle u^2\rangle-\langle u \rangle^2)/k_BT^2\propto L^{\gamma/\nu}.
\end{equation}
which scales asymptotically like the Ising {\em susceptibility} (fluid
compressibility). To recapture the Ising behaviour it is instead necessary 
to consider the fluctuations of the energy-like {\em operator}

\begin{equation}
C_v=L^d(\langle {\cal E}^2\rangle-\langle {\cal E} \rangle^2)/k_BT^2\propto L^{\alpha/\nu}.
\end{equation}

Further practical consequences of field mixing are to be found in the
finite-size behaviour of the average values of the critical density and
energy.  Specifically, recall that

\begin{equation}
\langle u \rangle_c=\langle\oE\rangle_c-\mixp\langle\oM\rangle_c
\hspace{1cm} \langle\rho\rangle_c=\langle\oM\rangle_c-\mix\langle\oE\rangle_c.
\label{eq:en+rho}
\end{equation}
Now, on symmetry grounds, the value of $\langle\oM\rangle_c=\int \pLM d\oM$ is
independent of $L$.  However, no such symmetry condition pertains to \pLE ,
whose average value $\langle\oE\rangle_c=\int d\oE\pLE$ at criticality 
varies with system size like

\begin{equation}
\langle\oE\rangle_c^L-\langle\oE\rangle_c^\infty \sim L^{-(1-\alpha)/\nu},
\label{eq:densca}
\end{equation}
implying the same behaviour for the energy and particle densities:
\setcounter{abc}{1}
\begin{eqnarray}
\langle u \rangle^L_c-\langle u \rangle^\infty_c\sim L^{-(1-\alpha)/\nu}\\
\addtocounter{equation}{-1}
\addtocounter{abc}{1}
\langle\rho\rangle^L_c- \langle\rho\rangle^\infty_c \sim L^{-(1-\alpha)/\nu}.
\end{eqnarray}
\setcounter{abc}{0}
Thus, even with precise knowledge of the location of the critical
point, a single measurement cannot yield accurate estimates of the infinite
volume values of $\rho_c$ and $u_c$.  Instead these quantities must be
estimated by employing the above scaling relations to extrapolate to the
thermodynamic limit data from a number of different system sizes
\cite{WILDING3}. 

It should also be pointed out, that the finite-size scaling expression,
equation~\ref{eq:critlim}, is strictly only valid in the limit of large
$L$.  For smaller system sizes, one anticipates that corrections to
finite-size scaling associated with non-vanishing values of the {\em
irrelevant} scaling fields will become significant \cite{WEGNER}.  These
irrelevant fields take the form
$a_1\tau^\theta+a_2\tau^{2\theta}+\ldots$, where $\theta$ is the
universal correction to scaling exponent, whose value has been estimated
to be $\theta\approx 0.54\pm 3$ for the 3D Ising class \cite{NICKEL}. 
Incorporating the least irrelevant of these corrections into
equation~\ref{eq:critlim}, one obtains \cite{PRIVMAN}

\begin{equation}
p_L({\cal M} , {\cal E}) \simeq  \Lambda _{\cal M}^+ \Lambda _{{\cal E}}^+ \tilde{p}^\star_{{\cal M},{\cal E}} (\Lambda _{\cal M}^+ \delta
{\cal M} ,
\Lambda _{{\cal E}}^+ \delta {\cal E}, a_1L^{-\theta/\nu})
\label{eq:cts}
\end{equation}
As we shall see in section~\ref{sec:lj}, it is necessary to
take account of such corrections to scaling, if highly accurate estimates of the
critical parameters are desired. 

Finally in this section, we note that the above treatment has been developed
(primarily for reasons of clarity) in terms of pure fluids of the Ising
class.  More generally though, the critical phenomena for a given fluid
system of interest may be different.  Thus for example the critical
behaviour of a binary mixture may be Ising like, but one must allow in
general for critical fluctuations both in the composition and in the total
density.  In this case the two scaling fields $\tau$ and $h$ will comprise
linear combinations of the three physical field i.e the temperature, the
chemical potential of one species, and the difference in chemical potential
between the two species \cite{SAAM}.  Alternatively one may encounter a
different universality class altogether, which may have more than just two
relevant scaling fields.  Examples are ternary mixtures of fluids or spin
fluids (like that considered in section~\ref{sec:tricrit}) which, by virtue
of their coupled order-parameters, exhibit tricritical phenomena.  In such
cases the tricritical behaviour is characterised by three or more scaling
fields, each of which comprises (in general) linear combinations of the
physical fields.  The nature and number of these physical fields will, of
course, depend upon the specific system under consideration. 

\section{The relationship to experiment}.

\label{sec:expt}

Experimental studies of critical phenomena in fluids have a long and
distinguished history, which it would be impossible to review here in any
great detail.  Instead we refer the reader to the recent book by Anisimov
\cite{ANISIMOV1} for a comprehensive treatment.  Here we shall focus on just
one aspect of fluid criticality, namely the field mixing phenomenon.  In so
doing we shall try to emphasise that simulation can often achieve much more
than just mimicking experiment.  Specifically we argue that by {\em
exploiting}, rather than simply trying to minimise finite-size effects, one
can often circumvent the difficulties encountered experimentally. 

Although theoretical predictions concerning the existence of field mixing
were confirmed experimentally quite some time ago
\cite{SENGERS,ANISIMOV,LEYKOO}, it has always proved rather difficult to
obtain accurate experimental data for quantities such as the field mixing
parameters.  The chief problem is that of the weakness
of the field mixing signature in the experimentally measurable observables. 
By contrast, a simulation is rather less hamstrung because it permits direct
access to {\em all} physical observables of relevance---notably the
exact instantaneous values of the particle and energy density, the coupling
of which manifests the field mixing phenomenon.  Moreover, the finite-size
of the system, which oftens bedevils computer simulation studies of critical
phenomena, turns out to be a positive boon for field mixing studies.  For
finite-sized systems, field mixing contributes an $L$-dependent correction
to the limiting forms of $p_L(\rho)$ and $p_L(u)$ (cf. 
section~\ref{sec:back}).  The effect of decreasing the system size is thus
to {\em magnify} the size of the field mixing signature in these
distributions.  Of course the system size shouldn't be too small, or else
corrections to finite-size scaling will become significant.  In most cases,
though, it turns out to be possible to attain sufficiently large system
sizes so that corrections to scaling are minimised, while nevertheless
retaining clear field mixing effects in $p_L(\rho)$ and $p_L(u)$.  As a
result (and as we will demonstrate in the next section), it is in most cases
relatively easy to measure accurately the field mixing parameters for the
system of interest. 

The situation for experiments is more complicated \cite{ANISIMOV,ANISIMOV1}. 
There one essentially operates in the thermodynamic limit and typically has
access only to bare observables such as the time-averaged density, or
response functions such as the isothermal compressibility or isochoric
specific heat.  For the density, the appearance of the energy operator in
equation~\ref{eq:udenmix} leads to the celebrated singularity in the
coexistence diameter \cite{SENGERS}:

\begin{equation} 
\rho_d=\frac{1}{2}(\rho_{liq}+\rho_{gas})=\rho_c+a\tau + b\tau^{1-\alpha} 
\end{equation} 
where $a$ is a constant representing the `rectilinear diameter' and $b$ is
closely related to the field mixing parameter $s$.  Unfortunately, owing to
the smallness of the exponent $\alpha\approx 0.11$ for fluids of the Ising
universality class, the power of the singular term is close to that of the
leading regular term.  Unless extremely small reduced temperatures are
attained, the two terms are difficult to distinguish experimentally.  Further
complications derive from gravity-induced density gradients in the
experimental sample (which are magnified by the large near-critical compressibility), as
well as hydrodynamic effects and convection currents \cite{ANISIMOV}.  Steps
must be taken either to minimise these effects during the experiment, or to
correct for them in the data analysis.  Owing to these difficulties,
relatively few experiments
\cite{SENGERS,LEYKOO} have been able to convincingly identify the diameter
singularity at all, let alone obtain accurate values for the field mixing
parameter $s$.  However, it is doubtful whether a simulation would fare much
better in attempting to investigate field mixing in this manner, even
without the problems of spurious sample inhomogeneities. 

The situation with regard to obtaining field mixing properties from
experimental measurements of the response functions, is also far from
straightforward.  The generalised compressibility and specific heat are
defined with respect to the scaling field derivatives of the operators,
whose singular behaviour is:

\begin{eqnarray}
\label{eq:response}
\setcounter{abc}{1}
\chi_\tau &=\left (\frac {\partial {\cal M}}{\partial h} \right )_\tau &\sim
\tau^{-\gamma} \\
\addtocounter{equation}{-1}
\addtocounter{abc}{1}
\chi_h &=\left (\frac {\partial {\cal E}}{\partial \tau} \right )_h &\sim
\tau^{-\alpha},
\end{eqnarray}
\setcounter{abc}{0}
which are the scaling equivalents for fluids of the Ising susceptibility and specific heat
respectively.  For pure fluids, however, the response functions
measured in practice are the isothermal compressibility and isochoric
specific heat:
\begin{eqnarray}
\setcounter{abc}{1}
\kappa_\tau &=&\left (\frac {\partial \rho}{\partial P} \right )_T\\
\addtocounter{equation}{-1}
\addtocounter{abc}{1}
C_V &=&\left (\frac {\partial u}{\partial T} \right )_\rho
\end{eqnarray}
\setcounter{abc}{0}
Asymptotically, it can be shown that these exhibit the same
scaling behaviour as the generalised response functions of
equation~\ref{eq:response}.  The only residual effect of field mixing is to
be found in additional non-asymptotic (sub-dominant) contributions. 
Specifically, one finds \cite{ANISIMOV}:

\begin{eqnarray}
\setcounter{abc}{1}
\kappa_{\tau} &=& A_0\tau^{-\gamma}+A_1\tau^{-\alpha}+\cdots\\
\addtocounter{equation}{-1}
\addtocounter{abc}{1}
C_V &=& B_0\tau^{-\alpha}+B_1\tau^{\gamma-2\alpha}+\cdots
\label{eq:compress}
\end{eqnarray}
\setcounter{abc}{0}
In practice, though, it seems difficult to unambiguously separate the
asymptotic and non-asymptotic terms from one another and from standard
corrections to scaling in the experimental data.


\section{Simulation studies}
\label{sec:results}

In this section we illustrate the practical application of the finite-size
scaling concepts outlined in section~\ref{sec:back}.  We do so by reviewing
(in some detail) two recent simulation studies that have applied FSS
analyses to investigate fluid critical behaviour.  The first system
considered in subsection~\ref{sec:lj}, is the well known Lennard-Jones
fluid.  We demonstrate how measurements of the scaling operator
distributions can be analysed within the FSS framework to yield highly
accurate estimates for the limiting critical point parameters of this model. 
We then turn to a more complicated system, namely a two-dimensional spin
fluid exhibiting a tricritical point.  We show how the FSS concepts can be
extended to incorporate the added complexity of tricritical phenomena, and
detail simulation results for the model.  Finally in
section~\ref{sec:complex} we consider recent applications of FSS techniques
to simulation studies of critical polymer blends and solutions. 

We begin, however, by briefly discussing some issues relating to the choice of
simulation ensemble. 

\subsection{Remarks on the choice of simulation ensemble}
\label{sec:ensemble}

The benefits that accrue from a FSS analysis of criticality in fluids are to
a large extent contingent upon the choice of simulation ensemble to be used. 
Crucial to the viability of the study is a choice of ensemble that
adequately provides for the strong near-critical order-parameter
fluctuations.  For liquid-vapour transitions, which are principally
characterised by density fluctuations, Monte-Carlo simulations \cite{MC}
within the grand canonical ensemble (constant $\mu VT$-ensemble)
\cite{ALLEN} have proved a highly effective approach.  The strength of the
grand canonical ensemble (GCE) stems from the freedom of the {\em total}
particle number to fluctuate \cite{WILDING5}.  Consequently, the near-critical
density fluctuations are observable on the largest possible length scale,
namely that of the system itself.  In principle, though, it is also possible
to perform a FSS analysis in the canonical $NVT$-ensemble (in which the total
density is fixed), by studying density fluctuations within {\em sub-blocks} of
the total system \cite{ROVERE,ROVERE1}.  However as described in
references~\cite{ROVERE1,WILDING5}, this approach is plagued with practical
difficulties and seems to be computationally very intensive.  The results
emerging from such studies have, to date, not been of comparable quality to
those obtained from GCE studies.
   
Another alternative approach for dealing with density fluctuations, is to
employ a constant pressure ($NpT$) ensemble \cite{ALLEN}.  Here, the total
particle number $N$ is held constant, but the density can fluctuate by means of
volume transitions.  Although FSS usually rest upon the idea of comparing
the correlation length to the (fixed) linear size of the system, its use has
recently been extended to the $NpT$-ensemble \cite{WILDING6} by expressing
the scaling properties not in terms of powers of $L$, but in powers of $N$. 
However, while fully feasible as a method for studying liquid-vapour
criticality, this approach was found (for various technical reasons) to be
less efficient than use of the GCE for the study of simple atomic
fluids \cite{WILDING6}.  It should nevertheless prove beneficial
when dealing with systems such as polymer solutions or electrolyte models
for which the GCE particle insertion probability can be prohibitively small. 

For liquid-liquid transitions, the near-critical region is characterised by
strong concentration fluctuations.  In a simulation one typically considers
two species of particles ($A$ and $B$) and common practice is to maintain the
total particle number, $N=N_A+N_B$, constant.  The analogue of the GCE in this
case is the semi-grand canonical ensemble (SGCE), in which concentration
fluctuations are realised by means of Monte-Carlo moves that swop the identity
of particles $A\to B$ or $B\to A$, under the control of a parameter
representing the chemical potential difference between the two species.  The
SGCE approach has recently been successfully employed in conjunction with a
FSS analysis of the ordering operator distribution $p_L({\cal M})$, to study
critical phenomena in a symmetric mixtures of square-well particles
\cite{DEMIGUEL} and in the Widom-Rowlinson model \cite{SHEW}.  The approach
can also be extended to asymmetric mixtures (where field mixing effects must
be taken into consideration) and even to asymmetric polymer blends as
described in section~\ref{sec:complex}.

Most of the recent work on fluid phase coexistence has been performed using
the Gibbs Ensemble Monte-Carlo (GEMC) method.  This technique, details and
applications of which are reviewed in reference \cite{GIBBS}, is very useful
for mapping out the phase coexistence envelope well away from criticality. 
However, in its most commonly practiced form, the method obtains estimates
for the critical parameters by extrapolating a power law to coexistence
curve data obtained well away from the immediate vicinity of the critical
point.  Due to various pitfalls of this approach \cite{MON,RECHT}, it has
thus far not proved able to provide more than rough estimates of critical
point parameters.  It is also not clear how the GEMC method can be combined
with FSS techniques.  For these reasons we will not consider such studies
further here. 

\subsection{The Liquid-vapour critical point of the Lennard-Jones fluid}
\label{sec:lj}

The Lennard-Jones (LJ) fluid, is arguably the simplest model with the
credentials (notably the symmetry) of a realistic atomic fluid.  Its
interparticle potential takes the form:

\begin{equation}
\phi(r)=4\epsilon_{LJ}[(\sigma/r)^{12}-(\sigma/r)^6]\label{eq:LJdef}
\end{equation}
where $\epsilon_{LJ}$ is the well-depth (which also defines the reduced temperature
$T^\star\equiv k_BT/\epsilon_{LJ}$), and $\sigma$ is a scale parameter.

Simulations of the LJ system have recently been carried out using a Metropolis
algorithm \cite{MC} within the grand canonical ensemble \cite{WILDING4}.  As
is common practice, the Lennard-Jones potential was truncated at a cutoff
radius $r_c=2.5\sigma$, and the potential left unshifted.  Five system sizes
were studied corresponding to $L=mr_c$ with $m=3,4,5,6,7$. The observables
recorded in the course of the simulations, were the reduced particle density,
$\rho^\star =L^{-d}N\sigma^3$ and the dimensionless energy density
$u^\star=L^{-d}(4\epsilon_{LJ})^{-1}\Phi(\{{\bf r}\})\sigma^3$, where $\Phi(\{{\bf
r}\})=\sum_{i\le j,j}\phi(|{\bf r_i}-{\bf r_j}|)$.  The joint distribution
$p_L(\rho^\star,u^\star)$ was accumulated in the form of a histogram.  A
number of preliminary short runs were made for the $m=4$ system size in order
to locate the liquid-vapour coexistence curve using a previous simulation
estimate of the critical temperature \cite{PANAGIO3}.  This was achieved by
tuning the chemical potential $\mu^\star$ until the density distribution
exhibited a double peaked structure indicative of phase coexistence.
Histogram reweighting \cite{FERRENBERG} was then used in order to explore the
coexistence curve in the neighbourhood of this simulation temperature.  Such a
reweighting scheme, use of which is now standard practice in simulation studies of
critical phenomena, allows histograms of observables obtained at one given $T$
and $\mu$ to be reweighted to yield estimates corresponding to another
$T^\prime=T+\Delta T$ and $\mu^\prime=\mu+\Delta\mu$.

\begin{figure}[h]
\vspace*{0.5 in}
\setlength{\epsfxsize}{12cm}
\centerline{\mbox{\epsffile{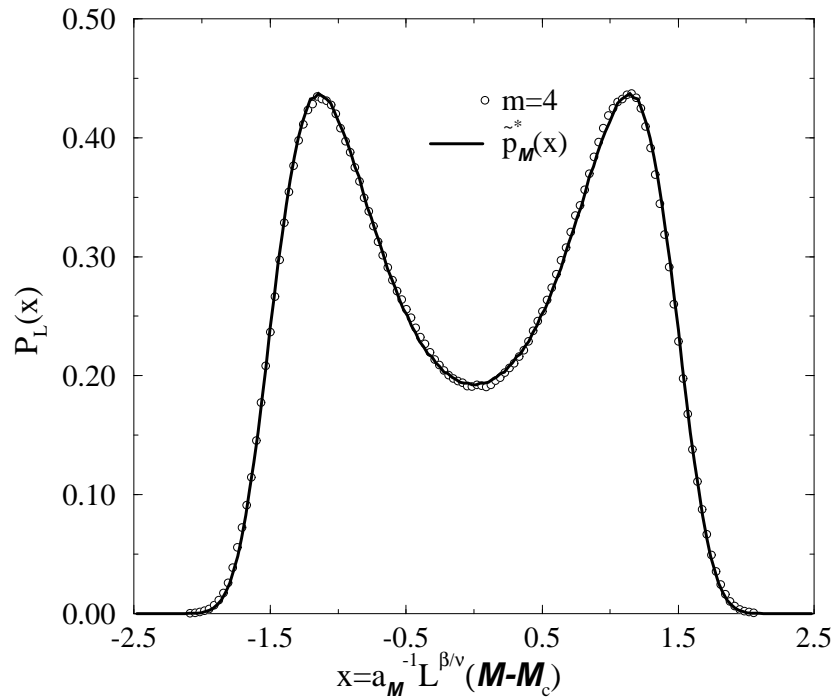}}} 
\vspace*{-0.2 in}
\caption{The measured form of the ordering operator distribution \pLM\
for the $m=4$ system size at the apparent critical parameters
$T_c^\star=1.1853,\mu_c^\star=-2.7843$. Also shown for comparison is
the universal fixed point ordering operator distribution \ptMstar
. The data has been expressed in terms of the scaling variable
$x=a_{\cal M}^{-1}L^{\beta/\nu}(\oM-\oM_c)$, with the value of the
 non-universal scale factor $a_{\cal M}^{-1}$  chosen so that the
distributions have unit variance. Statistical errors do not exceed the
symbol sizes. From ref. \protect\cite{WILDING4}}

\label{fig:oM_M4}
\end{figure}

To obtain a preliminary estimate of the critical point parameters, the
universal matching condition for the ordering operator distribution \pLM\
was invoked.  As observed in section~\ref{sec:back}, fluid--magnet
universality implies that the critical fluid ordering operator distribution
\pLM\ must match the universal fixed point function \ptMstar=$\int
dy$\ptMEstar\ $(x,y)$ appropriate to the Ising universality class.  This
latter function is independently known from detailed Ising model studies
\cite{HILFER}.  Thus the apparent critical parameters of the fluid can be estimated
by tuning the temperature, chemical potential and field mixing parameter $s$
(within the reweighting scheme) until \pLM\ collapses onto \ptMstar . 
The result of applying this procedure for the $m=4$ data set is displayed in
figure~\ref{fig:oM_M4} , where the data has been expressed in terms of the
scaling variable $x=a_{\cal M}^{-1}L^{\beta/\nu} (\oM-\oM_c)$ and in line
with convention, scaled to unit norm and variance.  The accord shown
corresponds to a choice of the apparent critical parameters
$T_c^\star(L)=1.1853, \mu_c^\star(L)=-2.7843$. 

Using this estimate of the critical point, more extensive simulations were
performed for each of the $5$ systems sizes, thus facilitating a full FSS
analysis.  Reweighting was again applied to the resulting histograms in order
to effect the matching of \pLM\ to \ptMstar , thus yielding values of the
apparent critical parameters.  Interestingly, however, the apparent critical
parameters determined in this manner were found to be $L$-{\em dependent}.
The reason for this turns out to be significant contributions to the measured
histograms from corrections to scaling (cf.  equation~\ref{eq:cts}), manifest
as an $L$-dependent discrepancy between the critical operator distributions
and their limiting fixed-point forms.  In the case of the ordering operator
distribution \pLM , the symmetry of the Ising problem implies that the
correction to scaling function is symmetric in $\oM - \langle \oM \rangle$.
In implementing the matching to \ptMstar , one thus necessarily introduces an {\em
additional} symmetric contribution to \pLM\ associated with a finite value of
the scaling field \gE .  This latter contribution has, coincidentally, a form
that is very similar to that of the correction to scaling function, a fact
that makes the cancellation of contributions possible.  It follows, therefore,
that the magnitude of the two contributions must be approximately equal.

\begin{figure}[h]
\vspace*{0.5 in}
\setlength{\epsfxsize}{12cm}
\centerline{\mbox{\epsffile{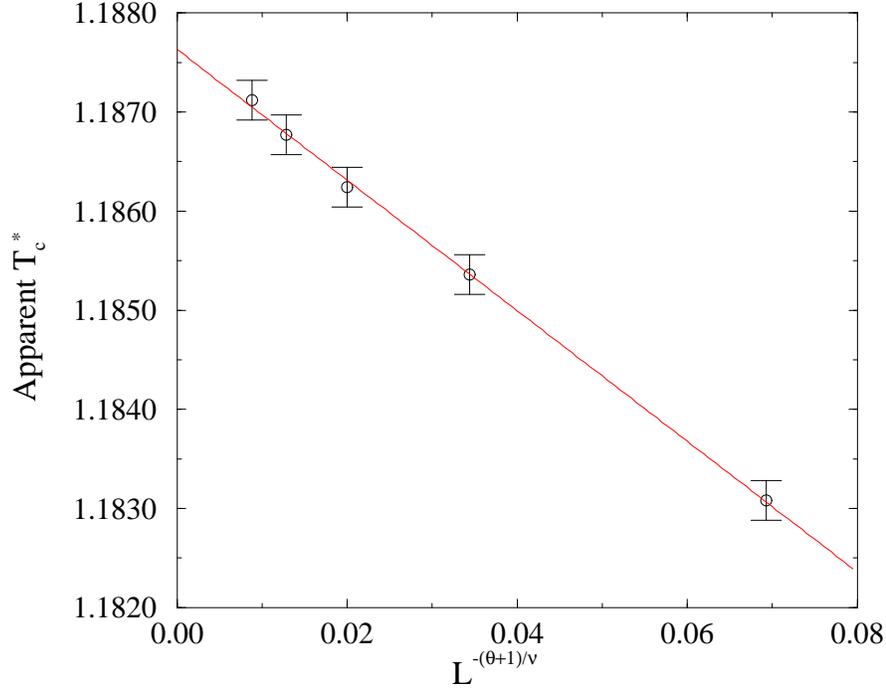}}} 

\caption{The apparent reduced critical temperature, (as defined by the
matching condition described in the text), plotted as a function of
$L^{-(\theta+1)/\nu}$, with $\theta=0.54$ and $\nu=0.629$
\protect\cite{NICKEL,FERRENBERG1}. The extrapolation of the least
squares fit to infinite volume yields the estimate
$T_c^\star=1.1876(3)$. From ref. \protect\cite{WILDING4}}

\label{fig:tc_extrap}
\end{figure}

Notwithstanding the complications engendered by corrections to scaling, it is
nevertheless possible to extract accurate estimates of the infinite-volume critical
parameters from the measured histograms.  The key to accomplishing this is
the known scaling behaviour of the corrections to scaling, which die away
with increasing system size like $L^{-\theta/\nu}$ (cf. 
equation~\ref{eq:cts}).  Since contributions to \pLM\ from finite values of
\gE , grow with system size like $\gE L^{1/\nu}$, it follows that
implementation of the matching condition leads to a deviation of the
apparent critical temperature $T_c^\star(L)$ from the true critical
temperature $T_c^\star$ which behaves like

\begin{equation}
T_c^\star(\infty) - T_c^\star(L) \propto L^{-(\theta+1)/\nu}.
\end{equation}
In figure~\ref{fig:tc_extrap} the apparent critical
temperature $T_c^\star (L)$ is plotted as a function of
$L^{-(\theta+1)/\nu}$. Clearly the data are indeed very well
described by a linear dependence, the least squares extrapolation of
which yields the infinite-volume estimate $T_c^\star=1.1876(3)$. The
corresponding estimate for the critical chemical potential is
$\mu_c^\star=-2.778(2)$.

\begin{figure}[h]
\vspace*{-0.3 in}
\setlength{\epsfxsize}{9cm}
\centerline{\mbox{\epsffile{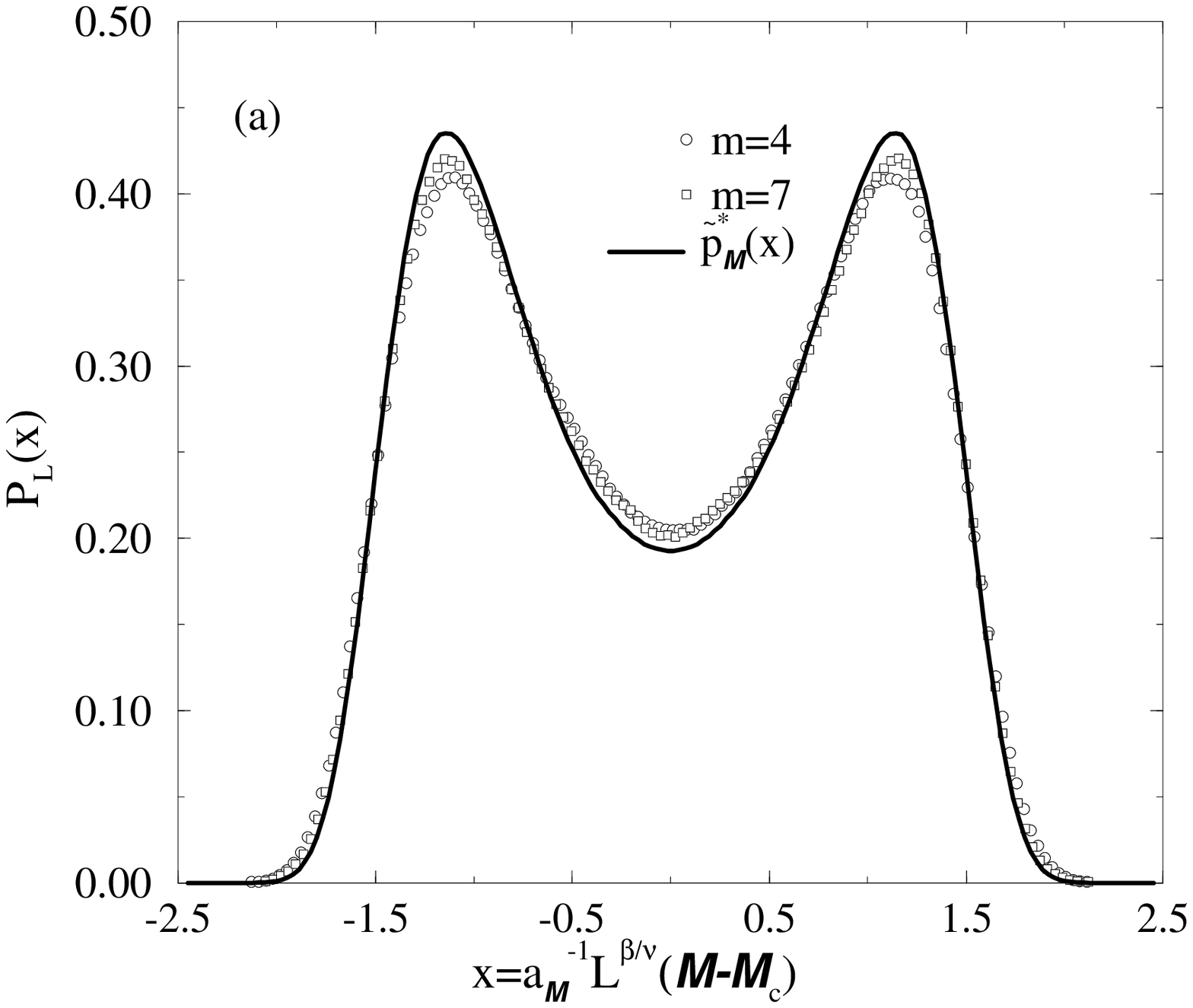}}} 
\vspace*{-0.3 in}
\centerline{\mbox{\epsffile{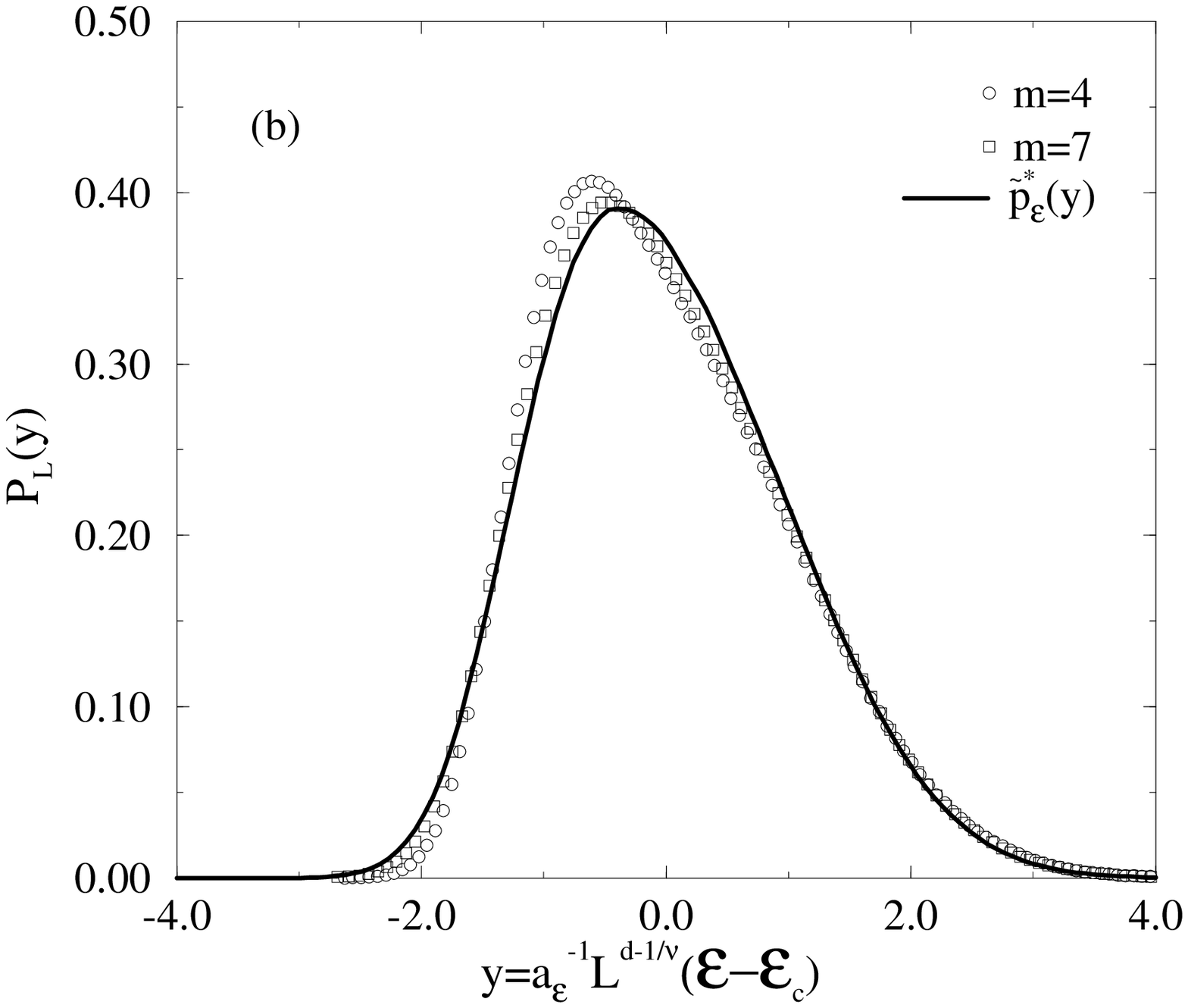}}} 

\caption{{\bf (a)} The ordering operator distribution $p_L({\cal M})$ for
the two system sizes $m=4$ and $m=7$ at the assigned critical parameters
$T_c^\star , \mu_c^\star$, expressed as function of the scaling variable
$x=a_{\cal M}^{-1}L^{\beta/\nu}({\cal M}-{\cal M}_c)$.  Also shown (solid
line) is the universal fixed point ordering operator distribution
$\tilde{p}^\star_{\cal M}(x)$.  {\bf (b)} The energy operator distribution
$p_L({\cal E})$ for the two system sizes $m=4$ and $m=7$ at $T_c^\star ,
\mu_c^\star$ ,expressed as a function of the scaling variable $y=a_{\cal
E}^{-1}L^{d-1/\nu}({\cal E}-{\cal E}_c)$.  Also shown (solid line) is the
universal fixed point energy operator distribution $\tilde{p}^\star_{\cal
E}(y)$ .  In both cases the values of the non-universal scale factors
$a_{\cal E}^{-1}$ or $a_{\cal M}^{-1}$ have been chosen to yield unit
variance.  From ref.  \protect\cite{WILDING4}}

\label{fig:oM+oE}
\end{figure}

The size and character of the contribution of corrections to scaling to the
operator distributions can be seen by plotting their forms at the estimated
infinite-volume values of $T_c^\star$ and $\mu_c^\star$.  Considering first
the ordering operator distribution, figure~\ref{fig:oM+oE}(a) shows the
critical point form of $p_L({\cal M})$ [expressed in terms of the scaling
variable $x=a_{\cal M}^{-1}L^{\beta/\nu}({\cal M}-{\cal M}_c)$], for the two
system sizes $m=4$ and $m=7$.  Also shown is the universal fixed point
function $\tilde{p}^\star_{\cal M}(x)$.  The corrections to scaling, manifest
in the discrepancy between the fluid finite-size data and the limiting form,
are clearly evident in the figure, especially for the $m=4$ system size.  The
corresponding situation for the energy operator distribution $p_L({\cal E})
=\int d {\cal M}p_L({\cal M},{\cal E})$ is shown in figure~\ref{fig:oM+oE}(b),
which is plotted together with the limiting fixed point function
$\tilde{p}^\star_{\cal E}(y)=\int
\tilde{p}^\star_{{\cal M},{\cal E}}(x,y) dx$ , identifiable as the critical
energy distribution of the Ising model, and independently known from detailed
Ising model studies \cite{WILDING3}.  The data have all been expressed in
terms of the scaling variable $y=a_{\cal E}^{-1}L^{d-1/\nu}({\cal E}-{\cal
E}_c)$.  In this case, one observes that the corrections to scaling are
noticeably larger than for $p_L({\cal M})$, a fact that presumably reflects
the relative weakness of critical fluctuations in ${\cal E}$ compared to those
in ${\cal M}$.

The values of the  critical point field mixing parameters, \mix\ and
\mixp, are implicit in the ordering operator distribution. Their values
may be estimated (cf. reference~\cite{WILDING3}) by treating each as a fit
parameter which is tuned to best optimise the mapping of the critical operator
distributions onto their limiting fixed point forms,
(cf. figure~\ref{fig:oM_M4}). The resulting estimates were, however, found to
be slightly $L$-dependent for the smaller system sizes, an effect that
presumably stems from corrections to scaling. For the two largest system
sizes, though, this $L$-dependence is small and the estimates $s=-0.11(1)$,
$r=-1.02(1)$ were obtained.

\begin{figure}[h]
\vspace*{-0.2 in}
\setlength{\epsfxsize}{11cm}
\centerline{\mbox{\epsffile{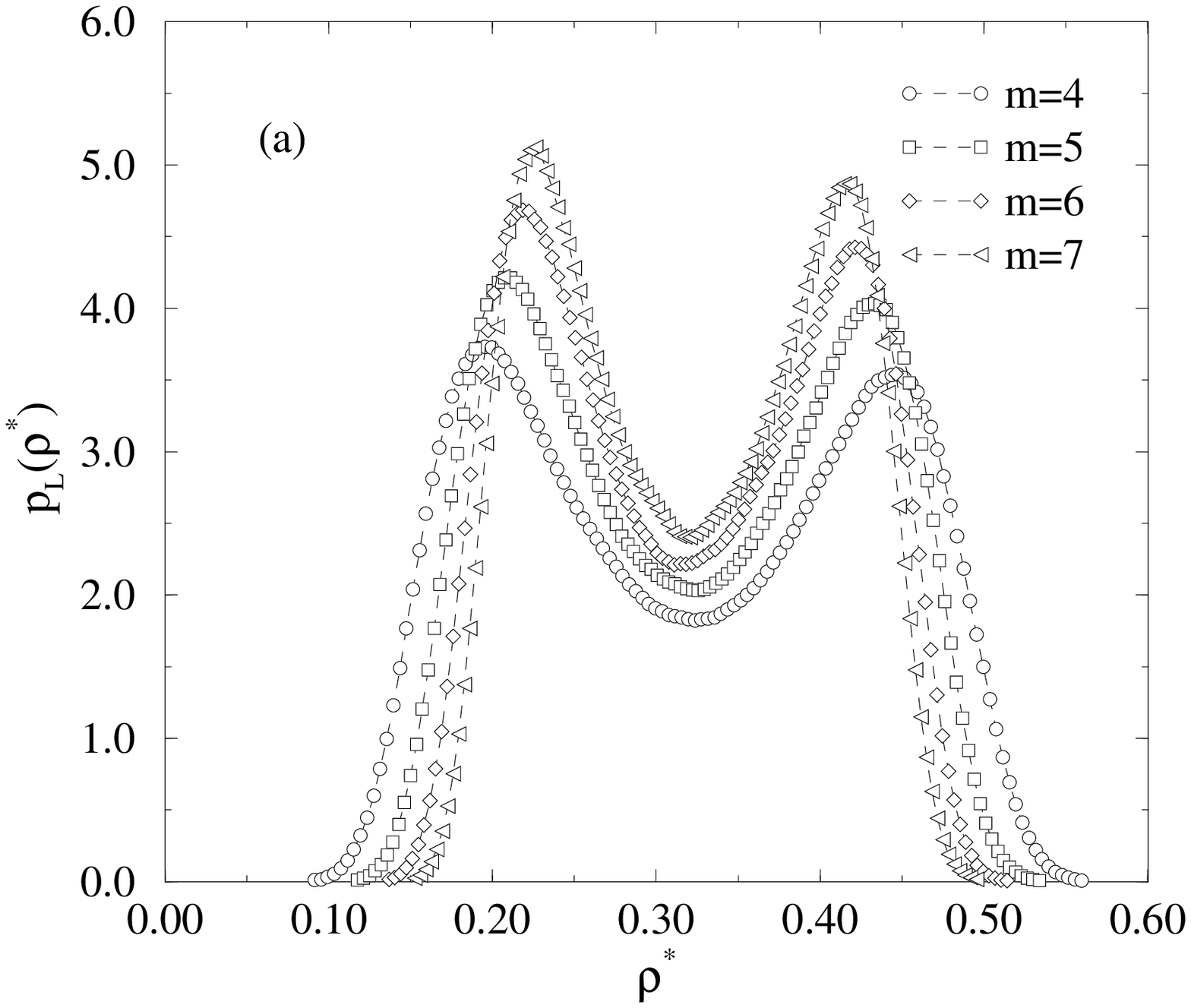}}}
\centerline{\mbox{\epsffile{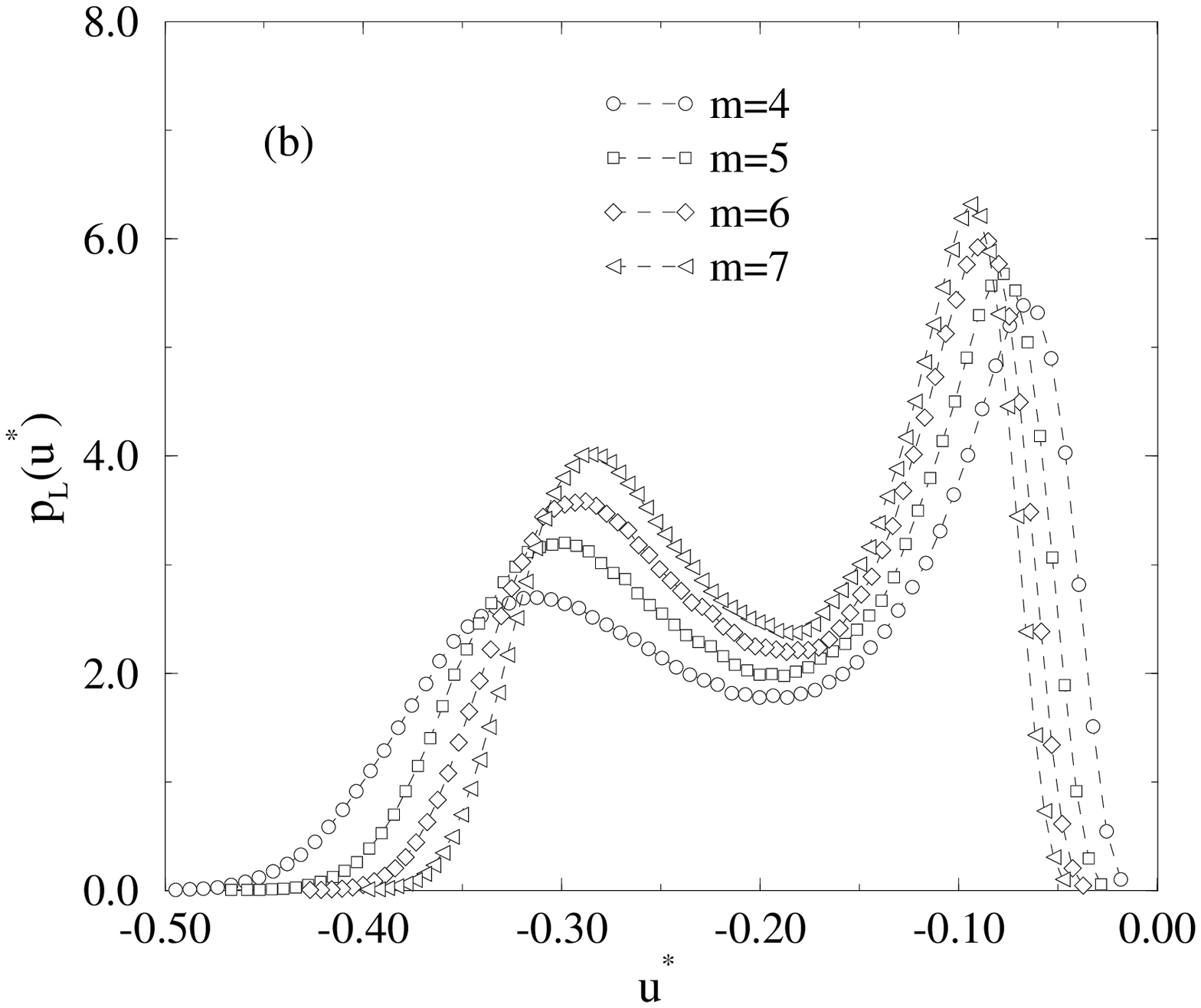}}}

\caption{{\bf (a)} The density distribution at $T_c^\star , \mu_c^\star$ for
the system sizes $m=4$---$7$, {\bf (b)} The corresponding energy density
distributions.  The lines are merely guides to the eye.  Statistical errors
do not exceed the symbol sizes.  From ref.  \protect\cite{WILDING4}}

\label{fig:rho+en}
\end{figure}

\begin{figure}[h]
\vspace*{-0.5 in}
\setlength{\epsfxsize}{11cm}
\centerline{\mbox{\epsffile{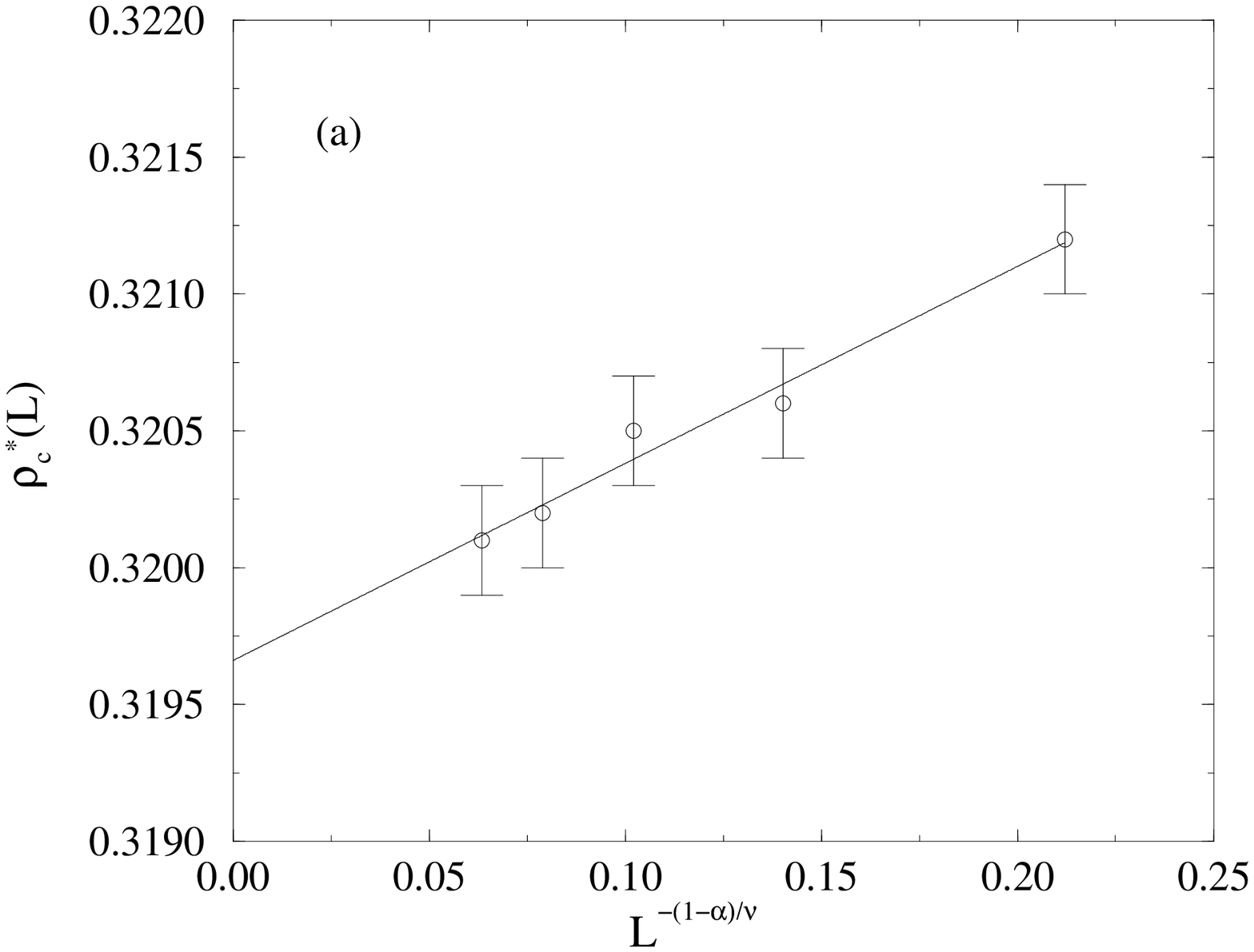}}}
\centerline{\mbox{\epsffile{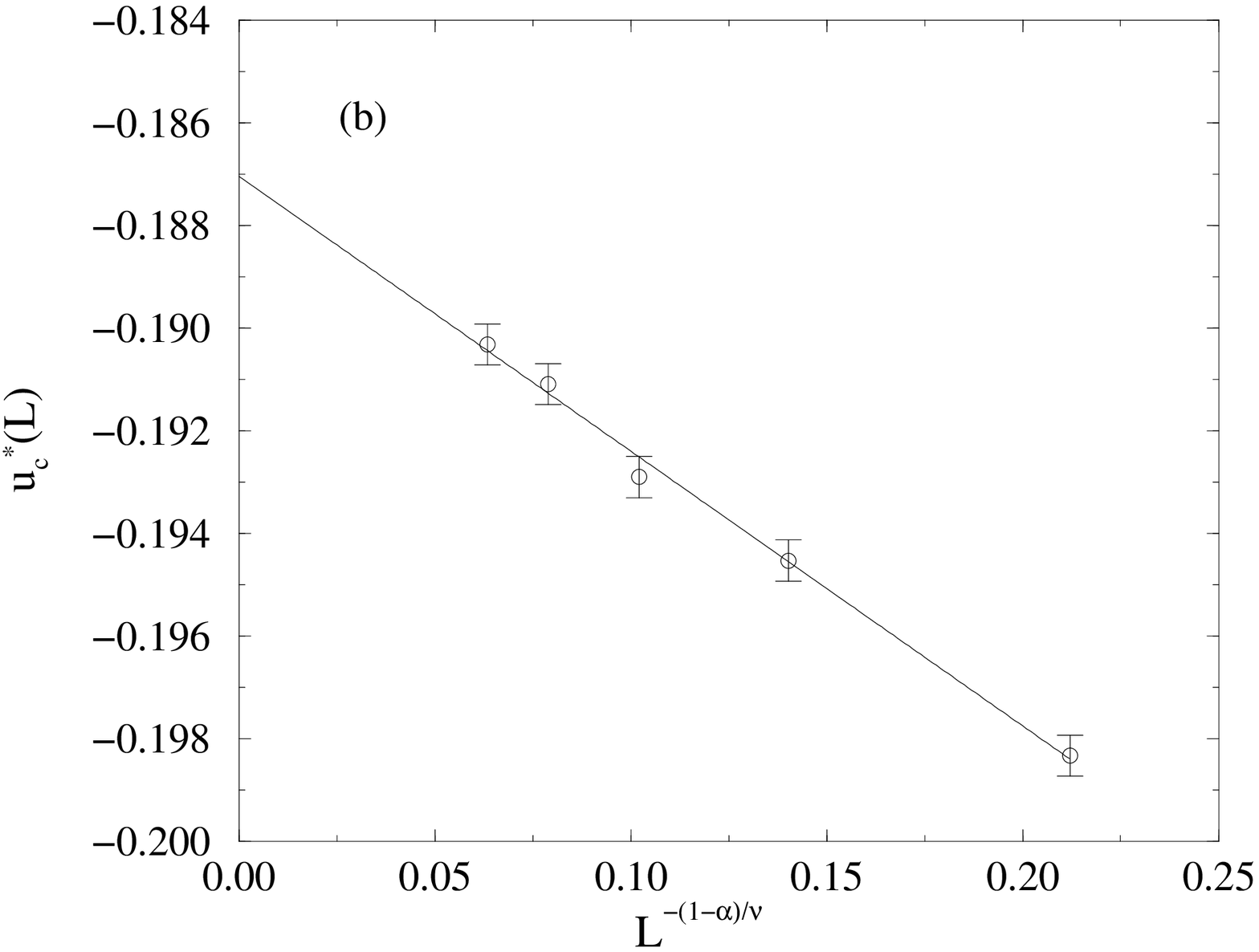}}} 

\caption{{\bf (a)} The measured average density
$\langle\rho\rangle_c(L)$ at the designated critical point, expressed
as a function of $L^{-(1-\alpha)/\nu}$. The least-squares fit yields
an infinite volume estimate $\rho_c=0.3197(4)$. {\bf (b)} The measured
average energy density $\langle u\rangle_c(L)$ at the critical point,
expressed as a function of $L^{-(1-\alpha)/\nu}$ . Extrapolation of
the least-squares fit to infinite volume yields the estimate
$u_c=-0.187(2)$. In both cases $1/\nu=1.5887$ was taken
\protect\cite{FERRENBERG1}. From ref. \protect\cite{WILDING4}}

\label{fig:rho+en_extrap}
\end{figure}

Addressing now the critical density and energy distributions,
figure~\ref{fig:rho+en} shows the measured forms of $p_L(\rho )$ and
$p_L(u)$ at the assigned values of the critical parameters.  To varying
degrees, these distributions are asymmetric, a fact which as explained in
section~\ref{sec:back} stems from field mixing effects.  Indeed, the forms
of the distributions are consistent with those shown in
figure~\ref{fig:projs}.  The magnitude of the asymmetry also clearly dies
away with increasing $L$, as predicted, although the system sizes are still
too small to reveal the limiting behaviour $p_L(\rho),p_L(u)\rightarrow
\ptMstar$.  In figure~\ref{fig:rho+en_extrap}, the values of $\langle \rho
\rangle_c$ and $\langle u \rangle_c$ corresponding to the distributions of
figure~\ref{fig:rho+en}, are plotted as a function of $L^{-(1-\alpha)/\nu}$
(c.f.  equation~\ref{eq:en+rho}).  Although no allowances have been made for
corrections to scaling (the effects of which are certainly much smaller than
those of field mixing), the data exhibit, within the uncertainties, a rather
clear linear dependence.  Least-squares fits to the data yield the infinite
volume estimates $\rho_c^\star= 0.3197(4), u_c^\star=-0.187(2)$. 

The finite-size scaling behaviour of $p_L({\cal M})$ and $p_L({\cal E})$ at
the critical point, also serve to furnish estimates of the exponent ratios
$\beta/\nu$ and $1/\nu$ characterising the two relevant scaling fields $h$
and $\tau$.  Consideration of the scaling form~\ref{eq:critlim} shows that
the typical size of the critical fluctuations in the energy-like operator
will vary with system size like $\delta{\cal E}\sim L^{-(d-1/\nu)}$, while
the typical size of the fluctuations in the ordering operator vary like
$\delta {\cal M}\sim L^{-\beta/\nu}$.  Comparison of the standard deviation
of these distributions as a function of system size thus affords estimates
of the appropriate exponent ratios.  In order to minimise systematic errors
arising from corrections to scaling, this comparison was performed only
for the two largest system sizes $m=6$ and $m=7$.  From the measured
variance of $p_L({\cal M})$ for these two systems, the estimate
$\beta/\nu=0.521(5)$ was obtained.  This compares most favourably with the
three dimensional (3D) Ising estimate \cite{FERRENBERG1} of
$\beta/\nu=0.518(7)$.  Given though that no allowances were made for
corrections to scaling, the quality of this accord is perhaps slightly
fortuitous. 

Implementing an analogous procedure for $p_L({\cal E})$, yields the estimate
$1/\nu=1.67(7)$, which does not agree to within error with the 3D Ising
estimate $1/\nu=1.5887(4)$.  In this case, however, it is likely that the
bulk of the discrepancy is traceable to the high sensitivity of $p_L({\cal
E})$ with respect to the designation of the field mixing parameter $r$
implicit in the definition of ${\cal E}$ (cf.  equation~\ref{eq:oplinks}). 
In the presence of sizable corrections to scaling, it is somewhat difficult
to gauge very accurately the infinite volume value of $r$ from the mapping
of $p_L({\cal E})$ onto $\tilde{p}^\star_{\cal E}(y)$.  Studies of
significantly larger system sizes than considered here would be necessary to
alleviate this problem. 

\subsection{Tricriticality in a two-dimensional spin fluid}
\label{sec:tricrit}

As the second example of fluid critical behaviour we consider the phase
behaviour of a two-dimensional (2D) `spin fluid', i.e.  a 2D fluid of
particles, each of which possesses a spin degree of freedom \cite{TRICRIT}. 
Physically, such a model is representative of a fluid of magnetic molecules
lightly absorbed onto a smooth substrate \cite{MARX}.  Owing to the coupling
of the magnetism and density fluctuations, however, the phase properties of
spin fluids are completely different from those of non-magnetic simple
fluids \cite{HEMMER,TAVARES} such as the LJ system. 

The simplest spin fluid model is a system of hard discs, each of
which carries a spin$-\frac{1}{2}$ magnetic moment.  Such a system has a
configurational energy given by:

\begin{equation}
\Phi(\{\vec{r},s\})=-\sum_{i<j}^NJ(r_{ij})s_is_j+\sum_{i<j}^NU(r_{ij})+H \sum_i
 s_i
\label{eq:pot}
\end{equation}
with $s_i=\pm 1$ , $U(r_{ij})$ is a hard disc potential
with diameter $\sigma$, and $H$ is an applied magnetic field.
The distant-dependent spin coupling parameter $J(r_{ij})$ is assigned 
a square well form:

\begin{eqnarray}
J(r)=\infty \hspace{1cm} & r < \sigma \nonumber \\
J(r)=J \hspace{1cm} &\sigma < r < 1.5 \sigma \nonumber \\
J(r)=0 \hspace{1cm} & r >1.5 \sigma \\
\end{eqnarray}

\begin{figure}[h]
\vspace*{0.1 in}
\setlength{\epsfxsize}{12cm}
\centerline{\mbox{\epsffile{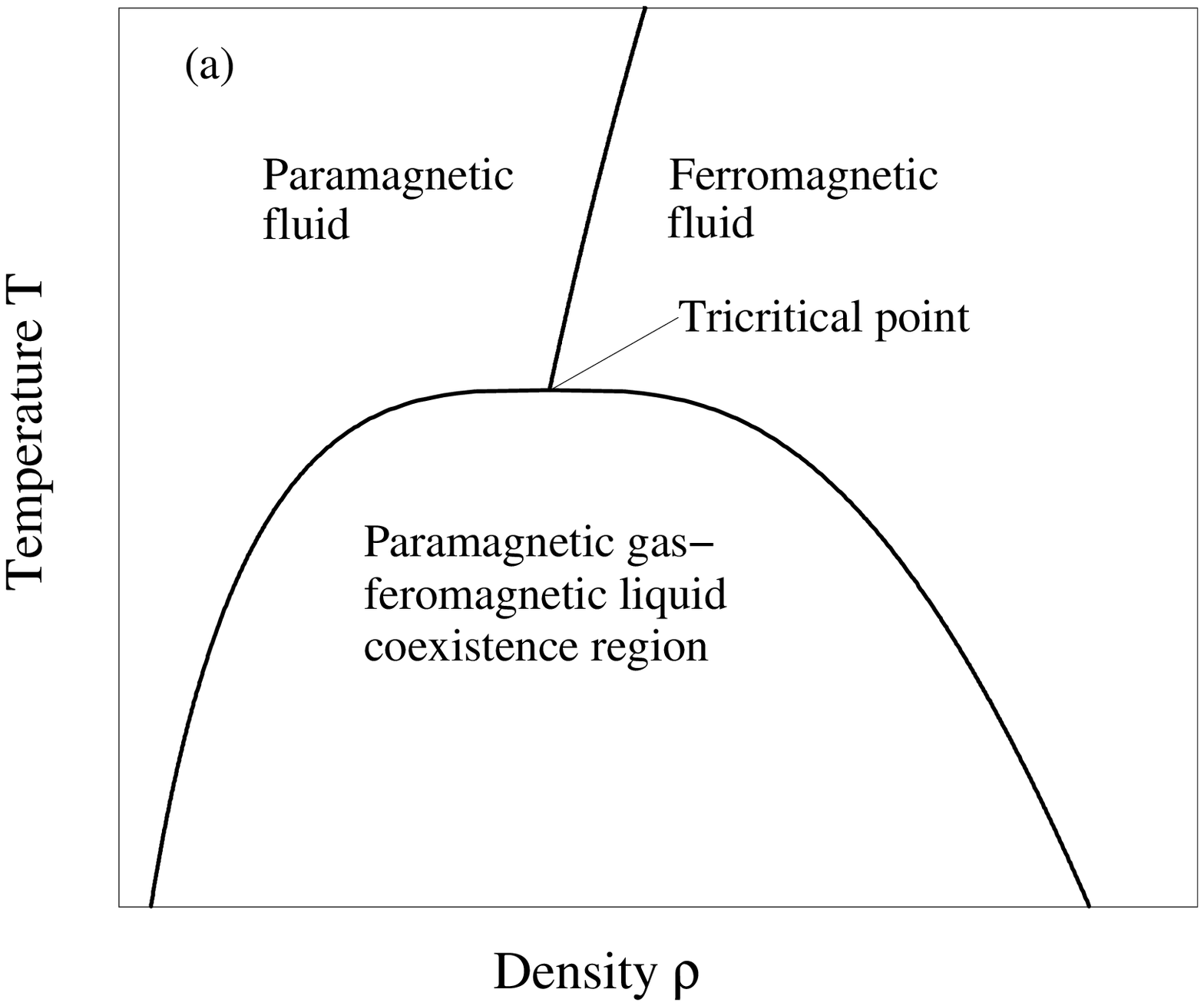}}} 
\centerline{\mbox{\epsffile{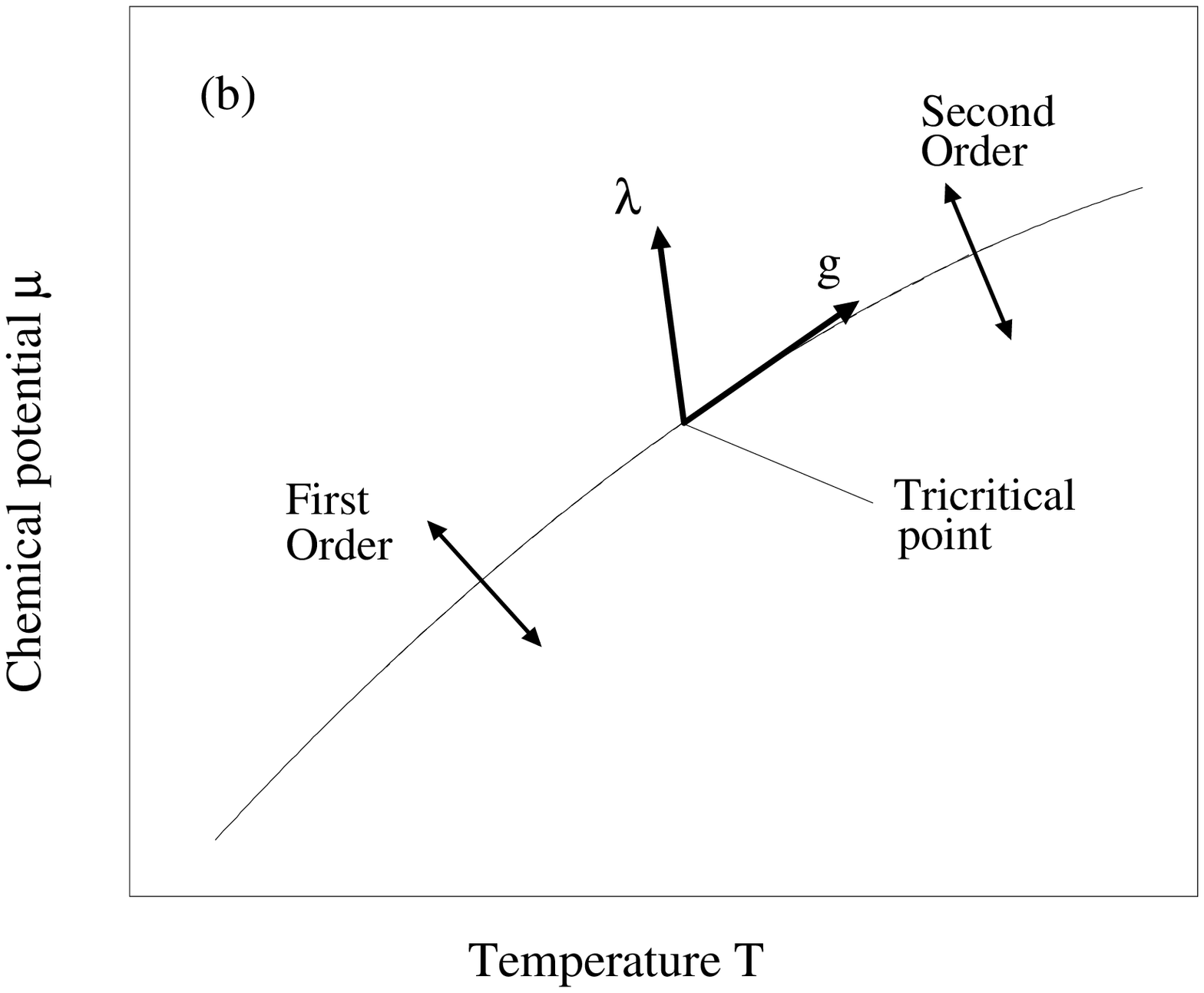}}} 
\caption{{\bf (a)} Schematic phase diagram of the spin fluid in the
$T$--$\rho$ plane. {\bf (b)} Schematic phase diagram in the $\mu$--$T$ plane
showing the directions of the relevant scaling fields $g$ and $\lambda$. 
From ref. \protect\cite{TRICRIT}}

\label{fig:pdschem}
\end{figure}

The schematic phase diagram of this model in zero field ($H=0$) is depicted in
figure~\ref{fig:pdschem}.  For high temperatures, there exists a line of
Ising critical points (the so-called `critical line') separating a
ferromagnetic fluid phase from a paramagnetic fluid phase.  The particle
density varies continuously across this line.  As one follows the critical
line to lower temperatures, the size of the particle density fluctuations
grows progressively.  At some point, the fluctuations in both the particle
density and magnetisation are simultaneously divergent and the system is
tricritical \cite{LAWRIE}.  Lowering the temperature still further results
in a phase separation between a low density paramagnetic gas and a high
density ferromagnetic liquid.  For subtricritical temperatures, the phase
transition between these two phases is first order. 

Owing to the interplay between the density and magnetisation fluctuations, the
tricritical properties of the spin fluid are expected to differ qualitatively
from those on the critical line.  General universality arguments
\cite{KADANOFF} predict that for a given spatial dimensionality, fluids with
short-ranged interactions should exhibit the same tricritical properties as
lattice-based spin systems such as the spin-$1$ Blume-Capel model
\cite{LAWRIE}.  However, since spin fluids possess a continuous translational
symmetry that lattice models do not, this proposal needs to be checked.  In
addition is if of interest to obtain the field mixing parameters of the model
and compare them with those for a simple critical fluid.

Before proceeding however, it is necessary to generalise somewhat the FSS
equations of section~\ref{sec:back}, in order to account for the fact that
the tricritical point has not two relevant scaling field, but three.  In
general these scaling fields (which we denote $g,\lambda$ and $h^\prime$)
are expected to comprise linear combinations of the three thermodynamic
fields $w\equiv J/k_bT$, $\mu$ and $H$.  For the spin fluid model considered
here, however, the configurational energy is invariant with respect to sign
reversal of the spin degrees of freedom and magnetic field.  This special
symmetry implies that the tricritical point lies in the symmetry plane
$H=0$, and that the scaling field $h^\prime$ coincides with the magnetic
field $H$, being orthogonal to the $\mu-w$ plane containing the other two
scaling fields, $g$ and $\lambda$.  Thus one can write

\begin{eqnarray}
\setcounter{abc}{1}
\label{eqn:scafldsa}
h^\prime & = & H-H_t \\ 
\addtocounter{abc}{1}
\addtocounter{equation}{-1}
\lambda & = & (\mu-\mu_t)+r(w_t-w) \\
\addtocounter{equation}{-1}
\addtocounter{abc}{1}
g & = & w_t-w+s(\mu-\mu_t) 
\label{eqn:scafldsc}
\end{eqnarray}
\setcounter{abc}{0}
where the subscript $t$ signifies tricritical values and the parameters $s$
and $r$ are field mixing parameters, controlling the directions of the scaling
fields in the $\mu$--$w$ plane.  The scaling fields $g$ and
$\lambda$ are depicted schematically in figure~\ref{fig:pdschem}(b). One sees
that $g$ is tangent to the coexistence curve at the tricritical point
\cite{GRIFF}, so that the field mixing parameter $r$ may be identified simply
as the limiting tricritical gradient of the coexistence curve. The scaling
field $\lambda$, on the other hand, is permitted to take a general direction
in the $\mu$--$w$ plane and is not constrained to coincide with any special
direction of the phase diagram \cite{REHR}.

Conjugate to each of the scaling fields are scaling operators, the
form of which are given by

\begin{eqnarray}
\setcounter{abc}{1}
{\cal M} & = & m \\ 
\addtocounter{abc}{1}
\addtocounter{equation}{-1}
{\cal D} & = & \frac{1}{1-sr}[\rho-su ] \\
\addtocounter{abc}{1}
\addtocounter{equation}{-1}
{\cal E} & = & \frac{1}{1-sr}[u-r\rho ] 
\end{eqnarray} 
\setcounter{abc}{0}
where $m$ is the magnetisation, $\rho$ is the particle density and $u$
is the energy density. In analogy to equation~\ref{eq:ansatz}, one can make the following
finite-size scaling {\em ansatz} for the limiting (large L)
near-tricritical distribution of $p_L(\rho,u,m)$

\begin{equation}
p_L(\rho,u,m)\simeq\frac{1}{1-sr}\tilde{p}_L(a_1^{-1}L^{d-y_1}{\cal M},
a_2^{-1}L^{d-y_2}{\cal D},a_3^{-1}L^{d-y_3}{\cal
E},a_1L^{y_1}h^\prime,a_2L^{y_2}\lambda,a_3L^{y_3}g)
\label{eq:ansatz_t}
\end{equation}
where $\tilde{p}_L$ is a universal scaling function, the $a_i$ are
non-universal metric factors and the $y_i$ are the standard tricritical
eigenvalue exponents \cite{GRIFF}, which in analogy to
equation~\ref{eq:ansatz} can be expressed in terms of ratios
of the various tricritical exponents characterising the behaviour of
$\rho$,$m$ and $\xi$ in the neighbourhood of the tricritical point. 

Precisely at the tricritical point, the tricritical scaling fields
vanish identically and the last three arguments of
equation~\ref{eq:ansatz_t} can be simply dropped, yielding 

\begin{equation}
p_L(\rho,u,m)\simeq\frac{1}{1-sr}\tilde{p}_L^\star(
a_1^{-1}L^{d-y_1}{\cal M},a_2^{-1}L^{d-y_2}{\cal D},a_3^{-1}L^{d-y_3}{\cal
E}),
\label{eq:ptri}
\end{equation}
where $\tilde{p}_L^\star$ is a universal and scale invariant function
characterising the tricritical fixed point.  In what follows we describe how the tricritical
point of the 2D spin fluid model was located by means of simulation, and how the
proposed universality of equation~\ref{eq:ptri} was tested by obtaining the
form of $\tilde{p}_L^\star$ and comparing it with that for the tricritical
2D Blume-Capel model whose Hamiltonian is given by

\begin{equation}
{\cal H}=-J\sum_{\langle ij\rangle }s_is_j+\Delta\sum_is_i^2, \hspace{5mm} s_i=-1,0,1
\end{equation} 
where $\Delta$ is the so-called crystal field (analogous to the chemical
potential). The phase diagram of the Blume-Capel model has the same topology
as that of figure~\ref{fig:pdschem}

Monte-Carlo simulations of the spin fluid model were performed using a
Metropolis algorithm within the grand canonical ensemble \cite{TRICRIT}. 
Three system sizes corresponding to $L=18\sigma, 24\sigma$ and $30\sigma$
were studied, containing (at liquid-vapour coexistence), average particle
numbers of $\langle N \rangle \approx 120$, $210$ and $330$ respectively. 
During the course of the simulation runs, the joint probability distribution
$p_L(\rho,u,m)$ was obtained in the form of a histogram.  The histogram
extrapolation technique \cite{FERRENBERG} was used to explore the
coexistence region. 

In contrast to the situation for the LJ fluid, no independent
knowledge of the tricritical scaling functions was available to aid
location of the tricritical parameters. It was thus necessary to
employ the cumulant intersection method to determine $T^\star_t$ and
$\mu^\star_t$ \cite{BINDER2}. The fourth order cumulant ratio $U_L$ is
a quantity that characterises the form of a distribution and is
defined in terms of the fourth and second moments of a given
distribution

\begin{equation}
U_L=1-\frac{<m^4>}{3<m^2>^2}.
\end{equation}
The tricritical scale invariance of the distributions $p_L({\cal D}),
p_L({\cal M})$ and $p_L({\cal E})$, (as expressed by
equation~\ref{eq:ptri}), implies that at the tricritical point (and
modulo corrections to scaling), the cumulant values for all system
sizes should be equal. The tricritical parameters can thus be found by
measuring $U_L$ for a number of temperatures and system sizes along
the first order line, according to the prescription outlined below.
Precisely at the tricritical temperature, the curves of $U_L$
corresponding to the various system sizes are expected to intersect
one another at a single common point.

\begin{figure}[h]
\vspace*{0.5 in}
\setlength{\epsfxsize}{14cm}
\centerline{\mbox{\epsffile{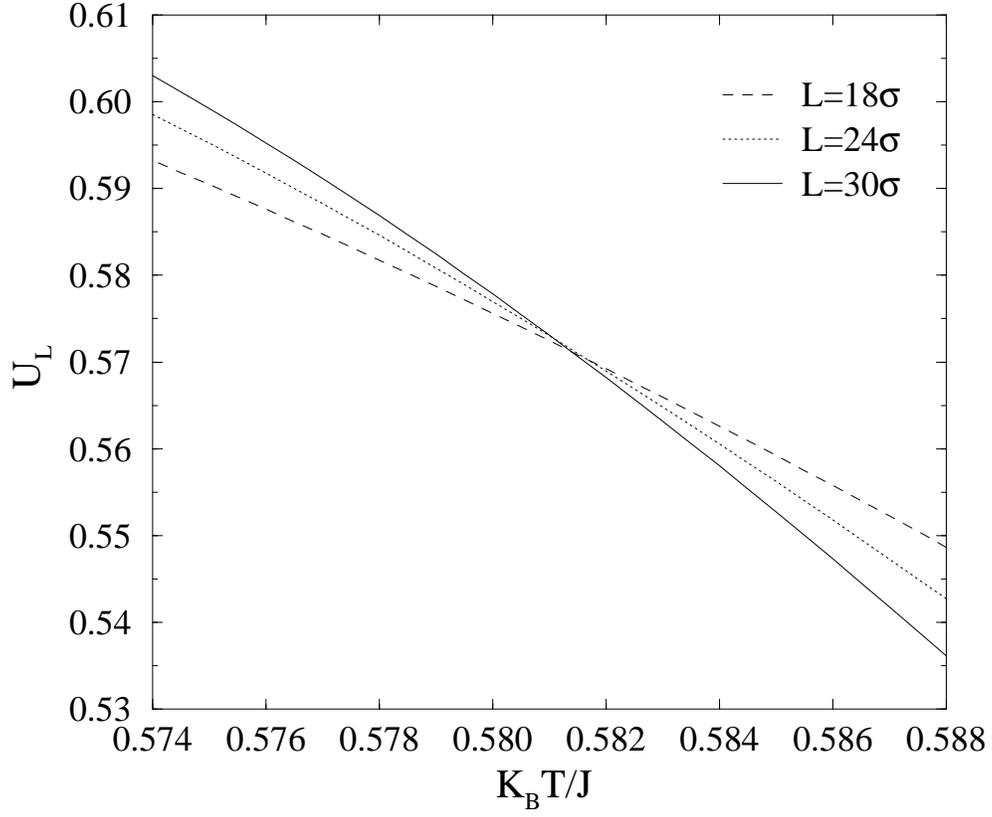}}} 

\caption{The measured cumulant ratios $U_L$ for the 2D spin fluid
model along the first order line and its analytic extension determined
according to the procedure described in the text. From ref. \protect\cite{TRICRIT}}

\label{fig:fl_UL}
\end{figure}

\begin{figure}[h]
\vspace*{0.5 in}
\setlength{\epsfxsize}{14cm}
\centerline{\mbox{\epsffile{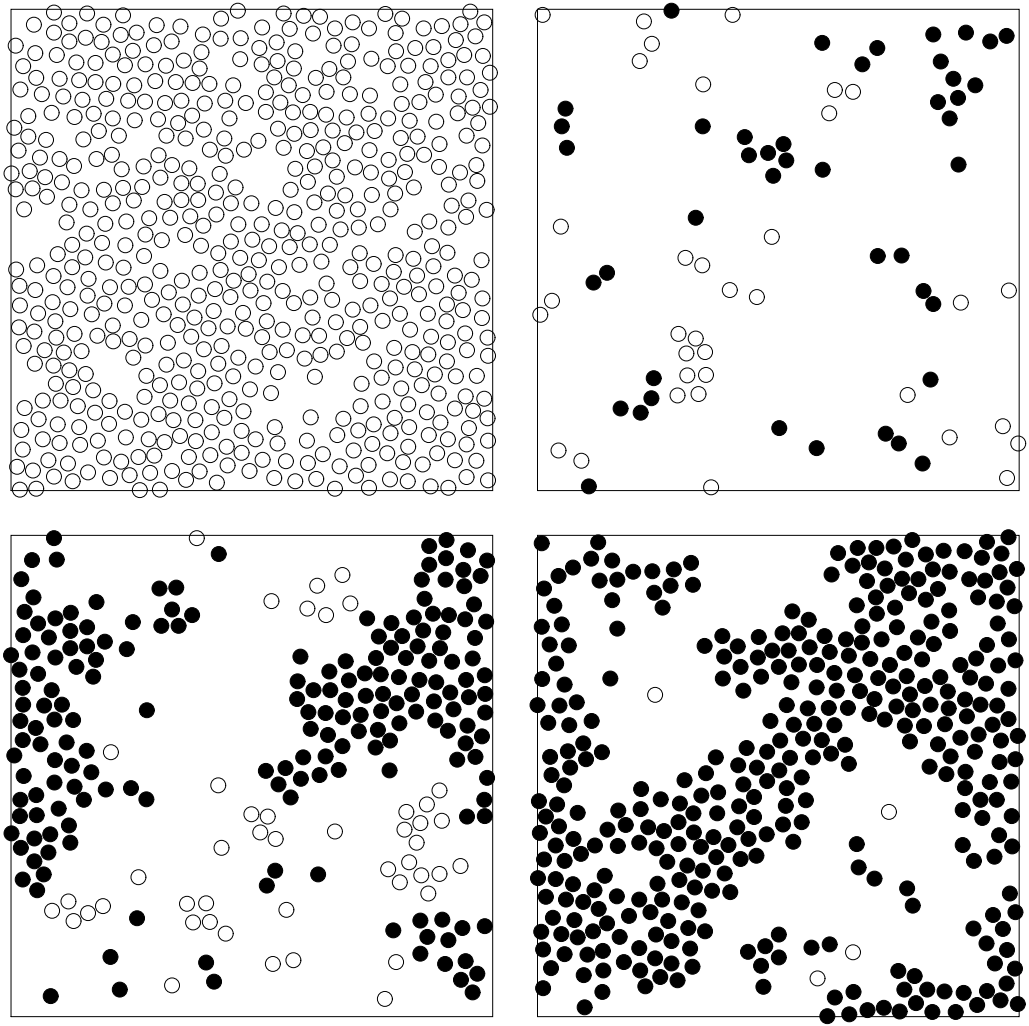}}} 

\caption{Typical particle/spin configurations of the $L=30\sigma$ spin fluid
near tricriticality. Spins values of $+1$ are denoted by filled circles,
and spin values $-1$ by unfilled circles. From ref. \protect\cite{TRICRIT}}

\label{fig:config}
\end{figure}

Initially, a very approximate estimate of the location of the tricritical
point was inferred from a number of short runs in which a temperature was
chosen and the chemical potential tuned while observing the density.  Long
runs were then carried out using this estimate for each of the three system
sizes.  From these runs the first order line and its analytic extension in
the $\mu$--$T$ plane were identified.  The measured values of $U^{\cal D}_L$
along this coexistence line are shown in figure~\ref{fig:fl_UL} for the each
system size.  Clearly the curves of figure~\ref{fig:fl_UL} have a single
well-defined intersection point, from which the tricritical temperature may
be estimated as being $k_BT_t/J=0.581(1)$.  The associated estimate for the
chemical potential is $\mu_t/k_BT=-1.916(2)$.  The tricritical particle and
energy densities were estimated to be $\rho_t=0.374(1)$ and $u_t=-0.778$
respectively.  The average magnetisation is of course strictly zero on
symmetry grounds.  Typical near-tricritical configurations for the
$L=30\sigma$ system are shown in figure ~\ref{fig:config}. They clearly show
the coupling of the density and magnetisation fluctuations, with a disordered
spins spin configuration at low density, and an ordered spin configuration
at high density.

\begin{figure}[h]
\vspace*{-1cm}
\setlength{\epsfxsize}{8.2cm}
\centerline{\mbox{\epsffile{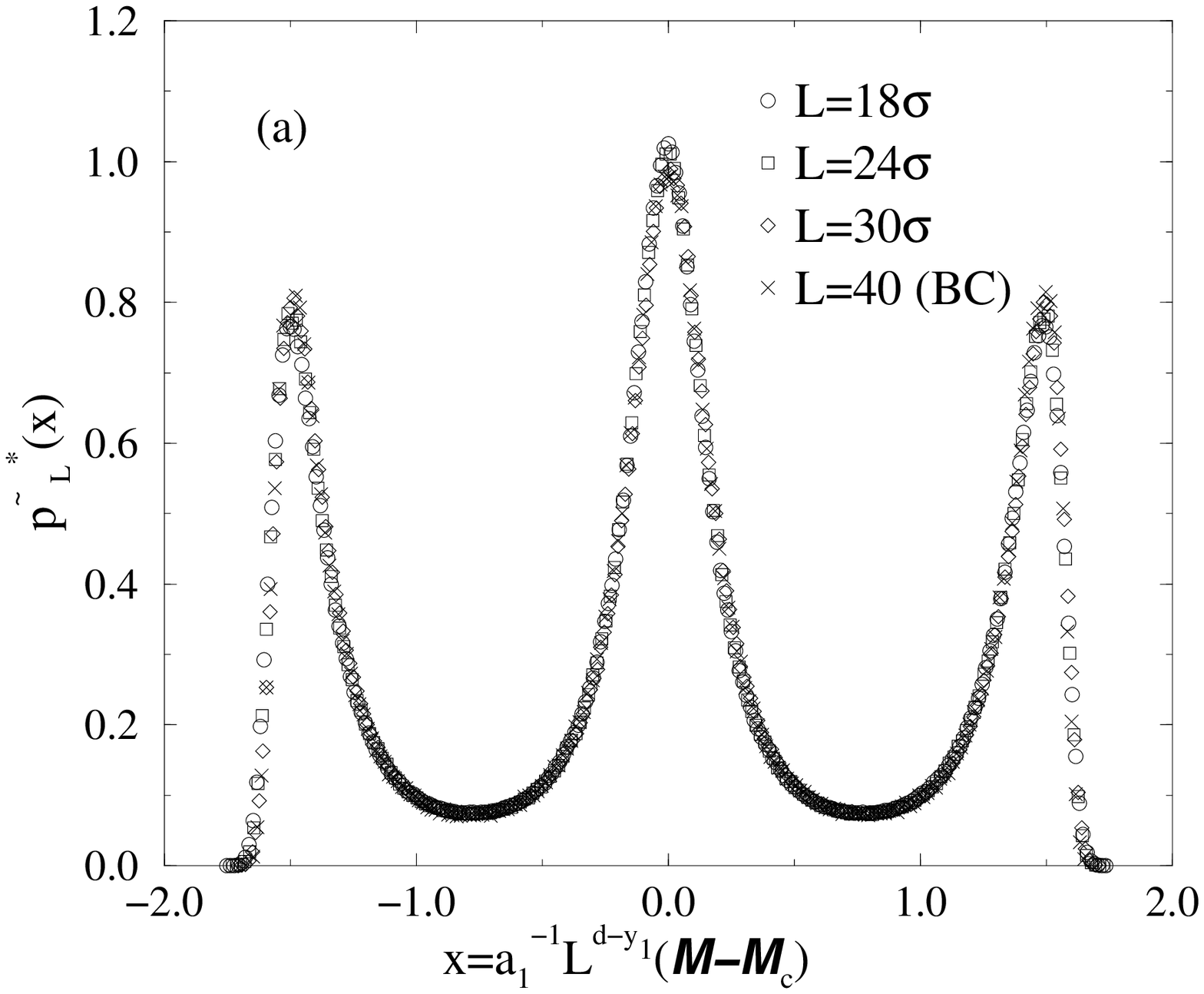}}} 
\centerline{\mbox{\epsffile{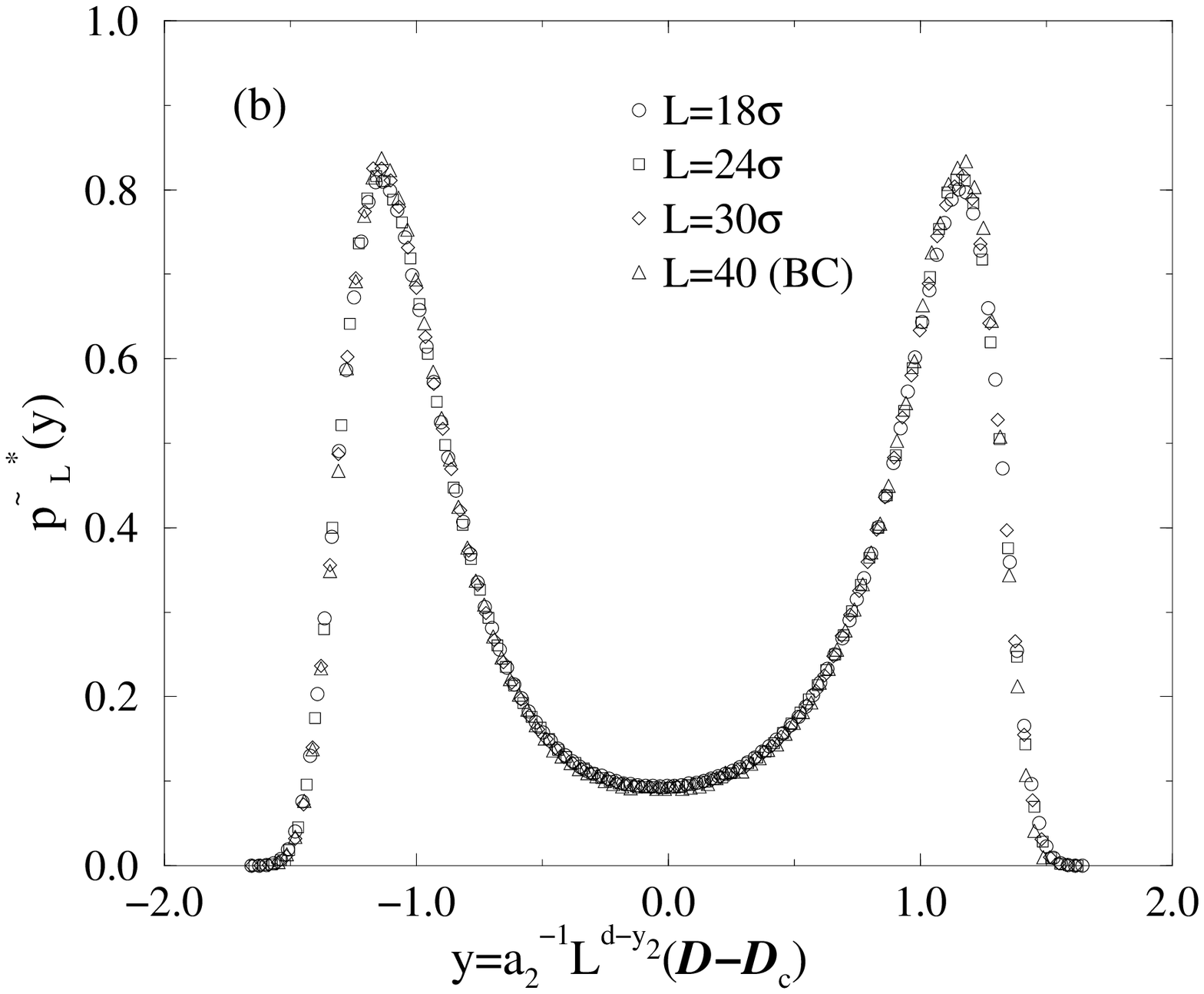}}} 
\centerline{\mbox{\epsffile{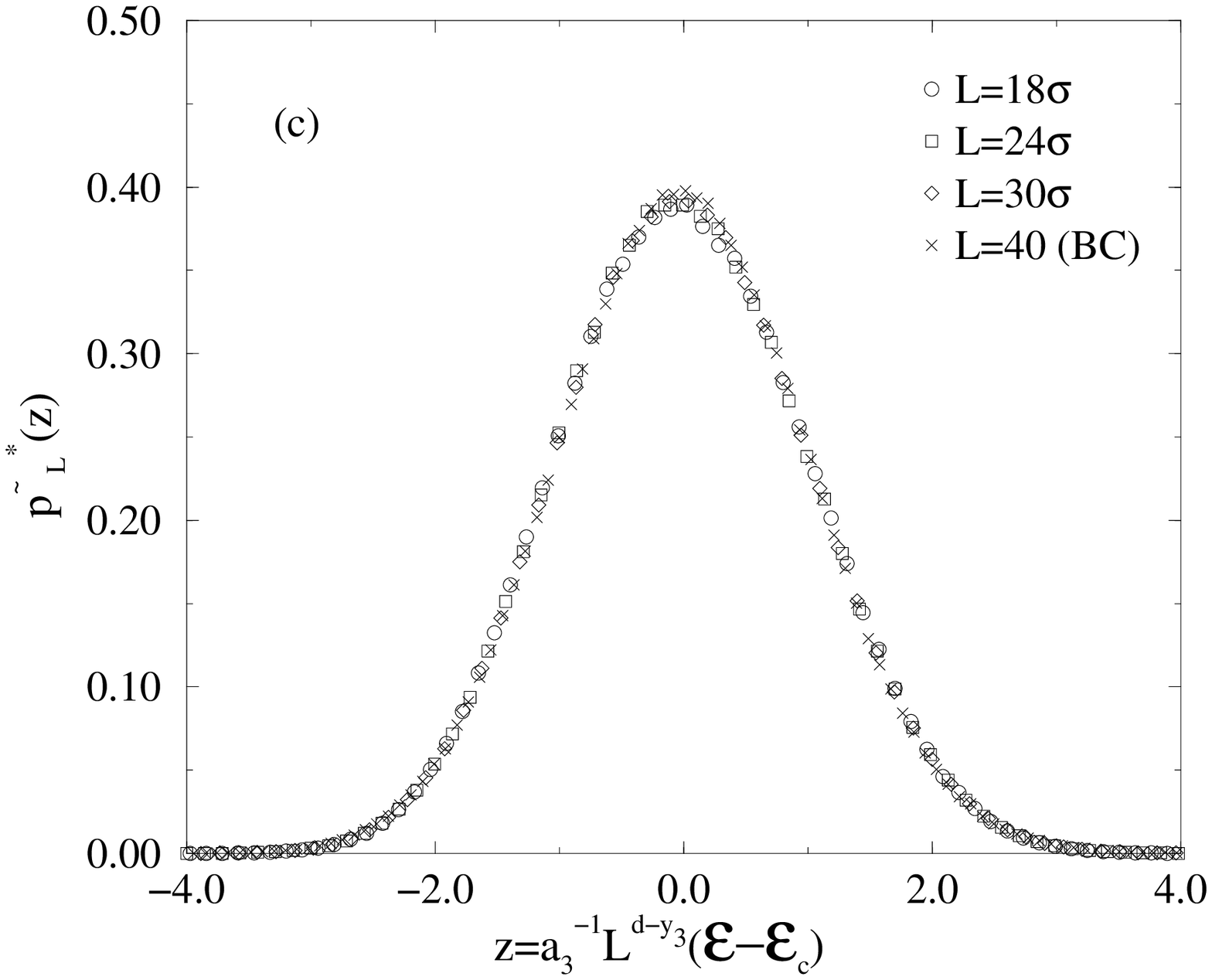}}} 

\caption{The scaling operator distributions for the 2D spin fluid at the
designated tricritical parameters for each of the three system sizes
$L=18\sigma, 24\sigma, 30\sigma$.  {\bf (a)} $\tilde{p}_L^\star({\cal M})$,
{\bf (b)}, $\tilde{p}_L^\star({\cal D})$ {\bf (c)} $\tilde{p}_L^\star({\cal
E})$.  Also shown for comparison are the corresponding distribution measured
for the tricritical $L=40$ 2D Blume-Capel model.  All distributions are
expressed in terms of the scaling variable $a_i^{-1}L^{d-y_i}({\cal O}-{\cal
O}_c)$ and are scaled to unit norm and variance.  Statistical errors do not
exceed the symbol sizes. From ref. \protect\cite{TRICRIT} }

\label{fig:collapse}
\end{figure}

Figure~\ref{fig:collapse} shows the forms of the operator distributions
$p_L({\cal M})$, $p_L({\cal D})$ and $p_L({\cal E})$ corresponding to the
designated values of the tricritical parameters.  Also included in
figure~\ref{fig:collapse} are the scaled tricritical operator distributions
obtained in a separate study of the 2D Blume-Capel model \cite{TRICRIT}. 
Clearly in each instance and for each system size, the scaled operator
distributions collapse extremely well onto one another as well as onto those
of the tricritical Blume-Capel model.  This data collapse is perhaps the
most stringent test of universality, and transcends critical exponent
values, although these too were obtained in the study of
reference~\cite{TRICRIT} and agree with those of the 2D Blume-Capel model. 
There can thus be little doubt that despite their very different microscopic
character, the spin fluid and Blume-Capel models do indeed share a common
fixed point. 

The values of the field mixing parameters $r$ and $s$, are implicit in the
forms of $p_L({\cal E})$ and $p_L({\cal D})$ shown in
figure~\ref{fig:collapse}, from which one finds \cite{TRICRIT} $r=-2.82$,
$s=-0.013$.  Intriguingly, this value of $s$ is at least an order of
magnitude smaller than that measured at the critical point of the 2D 
Lennard-Jones fluid \cite{WILDING1}.  This smallness implies that the
scaling field $\lambda$ almost coincides with the $\mu$ axis of the phase
diagram. 

With regard to the forms of the tricritical operator distributions, it is
found that $p^\star_L({\cal E})$ is (to within the precision of the
measurements) essentially Gaussian, implying that the tricritical
fluctuations in ${\cal E}$ are extremely weak.  The tricritical form of
$\tilde{p}_L^\star({\cal M})$ on the other hand is {\em three-peaked}, in
stark contrast to the situation in the 2D Ising model, where the critical
magnetisation distribution is strongly {\em double}-peaked
\cite{BINDER2,NICOLAIDES}.  This three-peaked structure reflects the
additional coupling between the magnetisation and the density fluctuations. 
Specifically, the central peak corresponds to fluctuation to small density,
which are accompanied by an overall reduction in the magnitude of the
magnetisation (cf.  figure~\ref{fig:config}).  Were one, however, to depart
from the tricritical point along the critical line, these density
fluctuations would gradually die out and a crossover to a magnetisation
distribution having the double-peaked Ising form would occur. 

Finally in this subsection, we mention that recent theoretical and
simulation studies have also been carried out on a 3D spin fluid model
having Heisenberg instead of Ising spins \cite{TAVARES,NIJMEIJER,MARCO}. 
GEMC simulations were carried out to determine the liquid-vapour phase
envelope, while the behaviour at points on the critical line of magnetic
transitions was investigated using a FSS analysis in the canonical ensemble. 
Interestingly the results were not in full accord with the known properties
of the lattice Heisenberg model, indicating a possible failure of
universality.  Further work is clearly called for in order to clarify this
issue. 

\subsection{Criticality in polymeric fluids}
\label{sec:complex}

Simulations of polymer systems are considerable more exacting in computational
terms than those of simple liquid or magnetic systems.  The difficulties stem
from the problems of dealing with the extended physical structure of polymers,
which gives rise to extremely slow diffusion rates, manifest in protracted
correlation times.  In order to ameliorate these difficulties, coarse-grained
lattice-based polymer models such as the bond fluctuation model (BFM)
\cite{BFM} are often employed.  Within the framework of the BFM
each monomer occupies a whole unit cell of a 3D periodic simple cubic lattice
and neighbouring monomers along the polymer chains are connected via one of
$108$ possible bond vectors, providing for $5$ different bond lengths and $87$
different bond angles. The model sacrifices chemically realistic detail for
greater computational tractability, while nevertheless retaining the essential
qualitative features of polymeric systems, namely chain connectivity and
excluded volume.  For studies of critical phenomena, this neglect of chemical
detail is not expected to bear on the universal scaling properties.

Most studies of criticality in polymeric systems have focussed on symmetric
blends i.e.  two species having identical chains length $N_A=N_B$.  SGCE
simulations (cf section~\ref{sec:ensemble}) of simple lattice walks were
studied by Sariban and Binder \cite{SARIBAN} who accurately located the
critical temperature by using the cumulant intersection method (cf. 
Section~\ref{sec:tricrit}).  A similar approach was applied by Deutsch and
Binder \cite{DEUTSCH1,DEUTSCH2} to the BFM, this time making extensive use
of histogram reweighting to map the phase diagram as a function of chain
length.  They confirmed the Ising character of the critical point and
reported tentative evidence for a crossover from Ising to mean field
behaviour away from the critical point \cite{DEUTSCH4}. 

Work on strongly asymmetric polymer blends $N_A\ne N_B$, is technically more
complicated than for symmetric mixtures, and has only become feasible very
recently.  The difficulties stem from excluded volume effects, which prevent
one simply removing a chain of one species and replacing it with another of
unequal size.  For fairly short chains ($N\le 30$) this problem can be
tackled using a novel simulation algorithm due to M\"{u}ller and Binder
\cite{MUELLER}, which operates by making MC moves that break an A-chain
into $k$ B-chains or joins $k$ B-chains to form a single A-chain.  This
algorithm was recently also employed by M\"{u}ller and Wilding
\cite{MUELLER1} to study the critical behaviour of a polymer blend within
the BFM, having $N_A=10, N_B=30$.  The observables sampled were the
concentration of $\rho_A$ of $A-$type chains (the concentration of $B$
chains being fixed by stipulating $N_A\rho_A+N_B\rho_B={\rm constant}$) and
the energy density.  FSS methods similar to those described in
sections~\ref{sec:lj} and ~\ref{sec:tricrit} were employed to accurately
locate the critical point and coexistence curve as a function of the
temperature and chemical potential difference between the two species.  The
Ising character of the critical point was clearly demonstrated and the field
mixing parameters obtained.  The scaling of the critical temperature with
chain length asymmetry was further studied using FSS by M\"{u}ller and
Binder~\cite{MUELLER2}, who found:

\begin{equation}
T_c\propto N_AN_B/(N_A^{1/2}+N_B^{1/2})^2,
\end{equation}
in agreement with Flory-Huggins mean-field theory. For short chains, however,
the mean field theory was found to  overestimate the critical temperature by about
$25\%$ due to the failure to treat properly the critical fluctuations.

Very recently, M\"{u}ller and Schick \cite{MUELLER3} have studied a ternary
polymer blend comprising two incompatible homopolymers and a symmetric
diblock copolymer.  Such a system has a phase diagram somewhat similar
to that of the spin fluid considered in section~\ref{sec:tricrit}, with
a line of second order transitions (corresponding to criticality between
the two homopolymer species) ending in a tricritical point below which
there is three-phase coexistence between two homopolymer rich phases and
a spatial structured copolymer-rich phase.  The SGCE simulation approach
was again employed  and the Ising character of the critical line
confirmed using FSS methods similar to those described in the 
section~\ref{sec:lj}.  The approximate location of the tricritical point
was also determined, although the accessible range of chain lengths and
system sizes were not sufficiently large to allow an unambiguous
characterisation of its nature. 

Finally, studies of the polymer-solvent critical point have recently been
reported by Wilding {\em et al} \cite{WILDING7}.  A biased chain insertion
scheme was employed \cite{FRENKEL}, allowing a GCE algorithm to be
implemented for the BFM.  Chains up to length $N=60$ monomers were studied
and a basic FSS study carried out in order to determine the chain length
dependence of the critical temperature and volume fraction.  For each chain
length investigated, the critical point parameters were determined by
matching the ordering operator distribution function to its universal
fixed-point Ising form.  Histogram reweighting methods were employed to
increase the efficiency of this procedure.  The results indicate that the
scaling of the critical temperature with chain length is relatively well
described by Flory theory, i.e.  $\Theta-T_c\sim N^{-0.5}$, where $\Theta$
is the temperature at which the chains behave ideally.  The critical volume
fraction, on the other hand, was found to scale like $\phi_c\sim N^{-0.37}$,
in clear disagreement with the Flory theory prediction $\phi_c\sim
N^{-0.5}$, but in good agreement with experiment \cite{WIDOM}.  Measurements
of the chain length dependence of the end-to-end distance suggested that the
chains are not collapsed at the critical point. 

\section{Discussion and Outlook}

In this review we have attempted to show that the finite-size scaling
techniques previously deployed with such success for lattice spin models,
can also be extended to permit accurate study of critical behaviour in
continuum fluids and polymeric systems.  In addition to illustrating how one
can tackle the critical region in practice, the applications reviewed were
intended to convey a flavour of the types of systems and phenomena that have
been studied.  Needless to say, however, there are a whole host of other
fluid systems which have not yet been studied in detail by simulation, and
whose critical phenomena raise a number of very interesting questions.  In
this section we attempt to highlight a small selection of them. 

One issue that is currently generating great theoretical and experimental
interest \cite{FISHER,FISHER1,LEVELT,EVANS,LEVIN}, is the thorny question of
the nature of criticality in ionic fluids.  Although the Ising character of
simple fluids with short range interactions is well established, the
theoretical understanding of ionic criticality is much less developed. 
Owing to their long ranged coulomb interactions, ionic fluids are not
expected {\em a-priori} to belong to the Ising universality class.  However
the possibility of an effective short-ranged interaction engendered by
charge-screening, does not preclude the possibility of asymptotic Ising
behaviour.  Indeed, experimentally, examples have been found of ionic fluids
exhibiting pure classical (mean-field) critical behaviour, pure Ising
behaviour, and in some cases a crossover from mean field behaviour to Ising
behaviour as the critical point is approached \cite{FISHER}.  It is thus of
considerable interest to determine the asymptotic universal behaviour, as
well as the physical factors controlling the size of the reduced crossover
temperature.  Simulation work on this problem to date has concentrated on
the prototype model for an ionic fluid, namely the restricted primitive
model (RPM) electrolyte, which comprises a fluid of hard spheres, half of which
carry a charge $+q$ and half of which carry a charge $-q$.  Unfortunately,
simulations of this model are hampered by slow equilibration and the need to
deal with the long range coulombic interactions \cite{MAKIS}.  Nevertheless
very recently, a FSS simulation study of the RPM has been reported, which
applied the techniques of section~\ref{sec:lj}, and which appears to show
that the critical point is indeed Ising like in character \cite{CAILLOL}. 
It remains to be seen though, to what extent this conclusion extends to
other less artificial electrolyte models, and whether simulation can shed
light on the system-specific factors controlling the disparaties in
crossover behaviour observed in real systems. 

Somewhat similar questions to those concerning ionic fluids, have also
recently been raised in the context of simple fluids with
variable interparticle interaction range.  Gibbs ensemble Monte Carlo
studies of the coexistence curve properties of the square-well fluid
\cite{VEGA} seem to suggest that while for short-ranged potentials the
shape of the coexistence curve is well described by an Ising critical exponent
$\beta\simeq 0.324$, the coexistence curve for longer ranged potentials is
near-parabolic in shape implying a classical (mean-field) exponent
$\beta\approx 0.5$.  This finding could doubtless be investigated in greater
detail using FSS techniques.  If a crossover does indeed occur, it would be
useful to try to formulate a Ginzburg criterion to describe its location as a
function of the interparticle interaction parameters. Experimental and
theoretical results concerning crossover from Ising to mean field behaviour in
insulating fluids and fluid mixtures have also recently been discussed in the
literature \cite{ANISIMOV2,ANISIMOV1}. 

Another matter of long-standing interest is the question of the physical
factors governing the size of the field mixing effect in fluids, which seem
to be strongly system specific.  Thus, for instance, the measured size of the
coexistence curve diameter singularity (cf.  subsection~\ref{sec:expt}) in
molten metals is typically much greater than that seen in insulating
fluids.  This has prompted the suggestion \cite{GOLDSTEIN} that the size of
the field mixing effect is mediated by the magnitude of three-body
interactions.  Clearly there is scope for simulation to attempt to
corroborate this proposal, possibly by introducing many body
interactions into a simple fluid model and observing the values of the field
mixing parameters as their strength is tuned. 

On a slightly different note to those emphasised in this review, we remark
that very little has been done to study interfacial effect close to the
critical point of realistic fluid models. Thus there are, to date, no
accurate simulation studies of the surface tension critical exponent or
associated amplitudes \cite{ZINN}. One might conceivably investigate this
matter by studying the temperature dependence of the ordering operator
distribution $p_L({\cal M})$ at coexistence, using preweighting techniques 
to overcome the free energy barrier, as has recently been done for the Ising
model and the Lennard-Jones fluid \cite{BERG,HUNTER,WILDING4}.

Turning finally to polymer systems, we note that recent advances in
simulation techniques, such as biased growth techniques and chain breaking
algorithms contribute significantly to our ability to deal with critical
phenomena in polymer systems.  Nevertheless, the studies reviewed in
subsection~\ref{sec:complex} were still limited to relatively small chain
lengths and system sizes and it is questionable whether the asymptotic
(large chain length) scaling properties are really being probed.  While
growing computational power will help somewhat to alleviate this problem,
further algorithmic improvements are clearly called for in order to deal
with even longer chains. 


\subsection*{Acknowledgements}

Much of the work reported here resulted from fruitful and enjoyable
collaborations with K.  Binder, A.D.  Bruce, M.  M\"{u}ller and P.  Nielaba. 
During some phases of the work the author received financial support from
the European Commission (ERB CHRX CT-930 351) and the Max-Planck Institut
f\"{u}r Polymerforschung, Mainz.

\begin{table}[h]
\begin{center}
\begin{tabular}{|c|c|c|} \hline
& Ising model & Pure fluids  \\ \hline
$\gE$ & $T-T_c$ & $w_c-w+s(\mu - \mu_c)$    \\
$\gM$ & $H-H_c$ & $\mu - \mu_c+ r(w_c-w)$     \\     
$\oE$ & $u$     & $\dfac \left[  u  - r \rho \right]$     \\     
$\oM$ & $m$     & $\dfac \left[ \rho - s u \right]$     \\ \hline
\end{tabular}
\end{center}
\caption{The forms of the relevant scaling fields and scaling
operators for the Ising magnet and for pure fluids.}
\label{table:relns}
\end{table}

\end{document}